\documentclass[secthm,seceqn,amsthm,ussrhead]{amsart}
\usepackage{amsmath,latexsym}
\usepackage[psamsfonts]{amssymb}
\usepackage{times}
\usepackage[mathcal]{euscript}
\numberwithin{equation}{section}
\newcommand{\bbR}{\mathbb R}
\newcommand{\bbT}{\mathbb T}
\newcommand{\bbZ}{\mathbb Z}

\renewcommand{\epsilon}{\varepsilon}

\newcommand{\be}{\begin{equation}}
\newcommand{\ee}{\end{equation}}

\newcommand{\F}{\mathbb{F}}

\newcommand{\N}{\mathbb{N}}

\newcommand{\R}{\mathbb{R}}

\newcommand{\T}{\mathbb{T}}

\newcommand{\Z}{\mathbb{Z}}

\newcommand{\cF}{{\mathcal F}}

\newcommand{\cH}{{\mathcal H}}

\newcommand{\cL}{{\mathcal L}}

\newcommand{\cU}{{\mathcal U}}

{\bf}{\it}
\newtheorem{theorem}{Theorem}[section]
\newtheorem{lemma}[theorem]{Lemma}
\newtheorem{corollary}[theorem]{Corollary}
\newtheorem{hypothesis}[theorem]{Hypothesis}
\newtheorem{definition}[theorem]{Definition}
\newtheorem{proposition}[theorem]{Proposition}
\newtheorem{remark}[theorem]{Remark}

\date{\today}

\begin{document}
\title{Schr\"{o}dinger
operators on lattices. The Efimov effect and discrete spectrum
asymptotics \\}

\author{Sergio	Albeverio$^{1,2,3}$, Saidakhmat	 N. Lakaev$^{4,5}$,
  Zahriddin I. Muminov$^{4}$}

\address{$^1$ Institut f\"{u}r Angewandte Mathematik,
Universit\"{a}t Bonn, Wegelerstr. 6, D-53115 Bonn\ (Germany)}

\address{
$^2$ \ SFB 256, \ Bonn, \ BiBoS, Bielefeld - Bonn;}
\address{
$^3$ \ CERFIM, Locarno and Acc.ARch,USI (Switzerland) E-mail
albeverio@uni.bonn.de}

\address{
{$^4$ Samarkand State University, Samarkand (Uzbekistan)} \
{E-mail: slakaev@mail.ru }}

\address{
{$^5$ Academy of Sciences of Uzbekistan}}

\address{
$^4$ {Samarkand State University, Samarkand (Uzbekistan)}{
E-mail:zmuminov@mail.ru}}

\begin{abstract}
The Hamiltonian of a system of three quantum mechanical particles
moving on the three-dimensional lattice $\Z^3$ and interacting via
zero-range attractive potentials is considered.
For the two-particle energy operator $h(k),$ with $k\in \T^3=(-\pi,\pi]^3$ the
two-particle quasi-momentum, the existence of a unique positive
eigenvalue below the bottom of the continuous spectrum of $h(k)$ for $k\neq0$
is proven, provided that $h(0)$ has a zero energy resonance.
The location of the essential and discrete spectra of the
three-particle discrete Schr\"{o}dinger operator $H(K),\,K\in \T^3$ being the
three-particle quasi-momentum, is studied. The existence of
infinitely many eigenvalues of $H(0)$ is proven.
It is found that for the number $N(0,z)$ of eigenvalues of $H(0)$ lying below
$z<0$ the following limit exists
$$
\lim _{z\to 0-} \frac {N(0,z)}{\mid \log\mid z\mid\mid }=\cU_0
$$
with $\cU_0>0$. Moreover, for all sufficiently small nonzero values of the
three-particle quasi-momentum $K$ the finiteness of the number $
N(K,\tau_{ess}(K))$ of eigenvalues of $H(K)$ below the essential spectrum
 is established and the asymptotics for the number $N(K,0)$
of eigenvalues	lying below zero is given.
\end{abstract}

\maketitle



\section{ Introduction}
	 One of the remarkable results in the spectral analysis for
continuous three-particle  Schr\"{o}dinger	operators  is the
Efimov	effect: if in a system of three-particles, interacting by
means of short-range pair potentials none of the three
two-particle subsystems has bound states with negative energy, but
at least two of them have a resonance with zero energy, then this
three-particle system has an infinite number of three-particle
bound states with negative energy, accumulating at zero.

This effect was first discovered by Efimov \cite{Ef}. Since then
this problem has been studied in many physics journals and books
\cite{AHW,AN,FM}. A rigorous mathematical proof of the existence
of Efimov's	 effect was originally carried out in [26] by Yafaev
and then in [21,23,24,25]. Efimov's effect was further studied in
[2,4,10,12,15,16,17,19,20].

Denote by $N(z),z<0$ the number of eigenvalues of the Hamiltonian
below $z<0$. The growth of $N(z)$ has been studied by
\medskip S. Albeverio, R. H\"{o}egh-Krohn, and T. T. Wu in [1] for
the symmetric case. Namely, the authors of [1] have first found
(without proofs) the exponential asymptotics of eigenvalues
corresponding to spherically symmetric bound states.

This result is consistent with the lower bound
$$
\lim_{z\rightarrow {0}} \inf \frac{N(z)}{|\log|z||}>0
$$
established in [24] without any symmetry assumptions.

The main result obtained by	 Sobolev [23] is the limit \be
{\lim_{z \rightarrow 0}} |\log |z||^{-1}N(z)={\cU}_0,
 \ee
where the coefficient ${\cU}_0$ does not depend on the potentials $%
v_\alpha $ and is a positive function of the ratios
$m_1/m_2,m_2/m_3$ of the masses of the three-particles.

In [2] the Fredholm determinant asymptotics of convolution
operators on large finite intervals with rational symbols having
real zeros are studied, as well as the connection with the Efimov
effect.

	In models of solid state physics [8,19,20,22], and also in
lattice	 field theory [9,18] discrete Schr\"{o}dinger operators
are considered, which are lattice analogs of the continuous
three-particle Schr\"{o}dinger operator. The presence of Efimov's
effect for these operators was demonstrated at the physical level
of rigor without a mathematical proof for a system of three
identical quantum particles in [19,20].

Although the energy operator of a system of three-particles on
lattice is bounded and the perturbation operator in the pair
problem is a  compact operator, the study of spectral properties
of energy operators of systems of two and three particles on a
lattice is more complex than in the continuous case.

In the continuous case [6] (see also [7,22]) the energy of the
center-of-mass motion can by separated out from the total
Hamiltonian, that is, the energy operator can by split into a sum
of a center-of-mass motion and a relative kinetic energy. So that
the three-particle "bound states " are eigenvectors of the
relative kinetic energy operator. Therefore Efimov's effect either
exists or does not exist for all values of the total momentum
simultaneously.

In lattice terms the "center-of-mass separation" corresponds to a
realization of the Hamiltonians as a "fibered operator", that is,
as the "direct integral of a family of operators" $H(K)$ depending
on the values of the total quasi-momentum $ K{\in}
{\bbT}^3=(-\pi,\pi]^3 $	 (see[8,22]). In this case a "bound state" is an eigenvector
of the operator $H(K)$ for some $ K {\in}{\bbT}^3 $. Typically,
this eigenvector depends continuously on K. Therefore, Efimov's
effect may exists only for some values of $ K{\in}{\bbT}^3$(see
[12]).

 In [10] was stated the existence infinitely many bound states (Efimov's effect)
for the discrete three-particle Schr\"{o}dinger operators
associated with a system of three arbitrary quantum particles
moving on three dimensional lattice and interacting via zero-range
attractive pairs potentials. In this work only a sketch of proof of
results has been given.

In [12] the existence of Efimov's effect for a system of three
identical quantum particles (bosons) on a three-dimensional
lattice interacting via zero-range attractive pair potentials has
been proven, in the case, where all three two-particle subsystems
have resonances at the bottom of the three-particle continuum.

In [13,14] the finiteness of the number of bound states was
proven, in the cases, where either none of the	two-particle
subsystems	or only one of the two-particle subsystems have a zero
energy resonance.

In [16] (a detailed proof is in [17]) the following results have
been established: for the difference operator on a lattice
associated with a system of three identical particles interacting
via zero-range attractive pair potentials  under the assumption
that all two-particle subsystems have resonance at the bottom of
the three-particle continuum.

1) for the zero value of the total quasi-momentum ($K=0$) there
are infinitely many eigenvalues lying below the bottom and
accumulating at the bottom of essential spectrum (Efimov's
effect).

2) for all $K \in U^{0}_{\delta}(0) =\{K\in {\bbT}^3:0<|K|<\delta
\}$, $\delta>0$ sufficiently small, the three-particle operator has
a finite number of eigenvalues below the bottom of essential
spectrum.

The results are quite surprising and clearly put in evidence the
difference between the continuum and discussed cases.

In the present work we consider a system of three arbitrary
quantum particles  on the three-dimensional lattice $\Z^3$
interacting via zero-range pair attractive potentials.

Let us denote by $\tau_{ess}(K)$ the bottom of essential spectrum
of the three-particle discrete Schr\"{o}dinger operator $H(K),\, K
\in {\bbT}^3$  and by $N(K,z)$ the number of eigenvalues lying
below
 $z \leq \tau_{ess}(K).$

The main results of the present paper are as follows:

(i) for the two-particle energy operator $h(k)$ on the
three-dimensional lattice $\bbZ^3$, $k$ being the two-particle
quasi-momentum, we prove the existence of a unique positive
eigenvalue below the bottom	 of the continuous spectrum of
$h(k),k\not=0$ for the nontrivial values of the quasi-momentum
$k$, provided that the two-particle Hamiltonian $h(0)$
corresponding to the zero value of $k$ has a zero energy
resonance.

(ii) we establish a location of the essential spectrum of the
 discrete three-particle operator $H(K)$. The infinitely many eigenvalues
 of the three-particle discrete Schr\"{o}dinger operator arise from the existence of resonances
 of the two-particle operators at the bottom of three-particle continuum.
Therefore we obtain a lower
 bound for the location of discrete
spectrum of $H(K)$ in terms of zero-range interaction potentials.

(iii) for the number $N(0,z)$ we obtain the limit result
$$
\lim_{z \to -0} \frac{N(0,z)}{|\log|z||}={\cU}_0,\, (0<{\cU}_0<
\infty).
$$

(iv) for any  $K \in U^{0}_{\delta}(0)$ we prove the finiteness of
$N(K,\tau_{ess}(K))$  and establish the following limit result
$$
\lim_{|K| \to 0} \frac{N(K,0)}{|\log|K||}=2{\cU}_0.
$$
We remark that whereas the result (iii) is similar to that of
continuous case and, the results (i) and (iv) are surprising and
characteristic for the lattice systems, in fact they do not have
any analogues in the continuous case.

The plan of	 the paper is as follows:

Section 1 is an introduction to whole paper.

In section 2 the Hamiltonians of systems of two and
three-particles in coordinate and momentum representations are
described as bounded self-adjoint operators in the corresponding
Hilbert spaces.

In section 3 we introduce the total quasi-momentum and decompose
the energy operators into von Neumann direct integrals, choosing
relative coordinate systems.

In section 4 we state the main results of the paper.

In section 5 we study spectral properties of the two-particle
discrete Schr\"{o}dinger operator $h(k),k\in \T^3$ on the
three-dimensional lattice $\bbZ^3.$ We prove the existence of
unique positive eigenvalue below the bottom of the continuous
spectrum of $h(k)$ (Theorem \ref{mavjud}) and obtain an
asymptotics for the Fredholm's determinant associated with $h(k).$

In section 6 we introduce the "channel operators" and describe its
spectrum by the spectrum of the two-particle discrete
Schr\"{o}dinger operators. Applying a Faddeev  type system of
integral equations we establish the location of the essential
spectrum (Theorem 4.3). We obtain a lower bound for the location
of discrete spectrum of $H(K)$ lying below the bottom of the
essential spectrum (see Theorem \ref{baho}). We receive a
generalization of the well known Birman -Schwinger principle for
the three-particle Schr\"{o}dinger operators on lattice (Theorem
\ref{g.b-s}) and  prove the finiteness of eigenvalues below the
bottom of the essential spectrum of $H(K)$ for $K\in
U^0_\delta(0)$ (Theorem \ref{finite}).

In section 7 we follow closely A.Sobolev method to derive the
asymptotics for the number of eigenvalues of the discrete
spectrum of $H(K)$ (Theorem \ref{asym.zK}).

Throughout the paper we adopt the following conventions: For each
$\delta>0$ the notation $U_{\delta}(0) =\{K\in {\bbT}^3:|K|<\delta
\}$ stands for a $\delta$-neighborhood of the origin and
$U^{0}_{\delta}(0)=U_{\delta}(0)\setminus\{0\}$ for a punctured
$\delta$-neighborhood. The subscript $\alpha$ (and also $\beta$
and $ \gamma$) always equal to $1$ or $2$ or $3$ and
$\alpha\neq\beta,\beta\neq\gamma,\gamma\neq\alpha.$

\section{Energy operators for two and three arbitrary
particles on a lattice in the coordinate and momentum
representations}

Let $\Z^{\nu}-\nu$ dimensional
 lattice. The free Hamiltonian $\widehat H_0$ of a system
of
 three quantum mechanical particles on the	three-dimensional
 lattice $\Z^3$ is defined in terms of three functions $
\hat{\varepsilon}_\alpha(\cdot)$ corresponding to the particles
$\alpha=1,2,3$ (called "dispersion functions" in the physical
literature, see,e.g. [19]).
 The operator $\widehat H_0$ usually associated with the following bounded
self-adjoint operator on the Hilbert space $\ell_2(({\bbZ}^3)^3)$:
\begin{equation*}
(\widehat{H}_0\hat{\psi})(x_1,x_2,x_3)=\sum_{s\in
{\Z}^3}[\hat{\varepsilon}_1(s) \hat{\psi}(x_1+s,x_2,x_3)+
\hat{\varepsilon}_2(s)\hat{\psi}(x_1,x_2+s,x_3)
\end{equation*}
\begin{equation*}
+\hat{\varepsilon}_3(s)\hat{\psi}(x_1,x_2,x_3+s)],\quad
\hat{\psi} \in \ell_2(({\bbZ}^3)^3).
\end{equation*}
Here $\hat{\varepsilon}_\alpha(\cdot), \alpha=1,2,3 $ are assumed
to be real-valued bounded functions of compact support on
 $\Z^3$ describing the dispersion low of the
 corresponding particles (see, e.g.,[19]).

The three-particle Hamiltonian $\widehat H$ of the
quantum-mechanical three-particles
 systems with two-particle pair	 interactions
$\hat v_{\beta\gamma},\beta,\gamma\in \{1,2,3\}$  is a bounded
perturbation of the free Hamiltonian $\widehat H_0$
\begin{equation}\label{total}
 \widehat{H}=\widehat{H}_0-\widehat{V}_{1}-\widehat{V}_{2}-
\widehat{V}_{3},
\end{equation}
where $\widehat{V}_{\alpha}, \alpha=1,2,3 $ are multiplication
operators on $\ell_2(({\bbZ}^3)^3)$
\begin{equation*}
(\widehat{V}_{\alpha}\hat{\psi})(x_1,x_2,x_3)=
\hat{v}_{\beta\gamma}(x_\beta-x_\gamma)\hat{\psi}(x_1,x_2,x_3),
\quad \hat{\psi} \in \ell_2(({\Z}^3)^3),
\end{equation*}
and $\hat{v}_{\beta\gamma}$ is bounded real-valued function.

Throughout this paper we assume that the following additional
Hypothesis holds.
\begin{hypothesis}\label{eps}
  The function $\hat{\varepsilon}_\alpha$ has the form
\begin{equation*}
 \hat{\varepsilon}_\alpha=l_\alpha \hat{\varepsilon},\quad
 \alpha=1,2,3,
 \end{equation*}
where $(l_\alpha)^{-1}>0,\,\alpha=1,2,3$ are different numbers,
having the meaning of mass of the particle $\alpha$;
 \begin{equation*}
 \hat{\varepsilon}:{\bbZ}^3\to {\bbR^1}
 \end{equation*}
 is given by
\begin{equation*}
\hat{\varepsilon}(s)=
 \begin{cases}
3, & s=0,\\
-\frac{1}{2}, & |s|=1,\\
0, & \text{otherwise}
 \end{cases}
 \end{equation*}
and $$\hat{v}_{\beta \gamma }(x_\beta-x_\gamma)
=\mu_{\alpha}\delta_{x_\beta x_\gamma},$$ where $\mu_\alpha>0$
interaction energy of particles $\beta$ and $\gamma$,
$\delta_{x_\beta x_\gamma}$ is Kroneker delta.
\end{hypothesis}

It is clear that under Hypothesis \ref{eps} the three-particle
 Hamiltonian \eqref{total} is  a bounded self-adjoint operator on the Hilbert space
 $\ell_2(({\bbZ}^3)^3)$.

Similarly as we introduced $\widehat H,$ we shall introduce the
corresponding two-particle Hamiltonians $\hat{h}_\alpha,\,\alpha
=1,2,3$ as bounded self-adjoint operators on the Hilbert space
$\ell_2(({\bbZ}^3)^2)$
\begin{equation*}
\hat{h}_\alpha =\hat{h}_\alpha^0- \hat{v}_{\alpha },
\end{equation*}
where
\begin{equation*}
(\hat{h}_\alpha^0\hat{\varphi})(x_\beta,x_\gamma) =\sum_{s\in {
\bbZ}^3}[\hat{ \varepsilon}_\beta
(s)\hat{\varphi}(x_\beta+s,x_\gamma)+\hat{\varepsilon}_\gamma (s)
\hat{\varphi}(x_\beta,x_\gamma+s)]
\end{equation*}
and
\begin{equation*}
(\hat{v}_{\alpha}\hat{\varphi})(x_\beta,x_\gamma)
=\mu_{\alpha}\delta_{x_\beta
x_\gamma}\hat{\varphi}(x_\beta,x_\gamma),\quad \hat{\varphi} \in
\ell_2(({\bbZ}^3)^2).
\end{equation*}

Let us rewrite our operators in the momentum representation.
 Let ${ \cF}_m:L_2(({\bbT}^3)^m) \rightarrow \ell_2((
{\bbZ}^3)^m)$ denote the standard Fourier transform,
 where ${({\T}^3)^m},\,m\in \N$ denotes the Cartesian
$m$-th power of the set ${\bbT}^3=(-\pi,\pi]^3.$

The three-resp. two-particle Hamiltonians (in the momentum
representation) are given by the bounded self-adjoint operators on
the Hilbert spaces
 $L_2(({\T}^3)^3)$ resp. $L_2(({\T}^3)^2)$
 as follows
 \begin{equation*}
 H={\cF}_3^{-1} \widehat H {\cF}_3
 \end{equation*}
resp.
 \begin{equation*}
h_\alpha={\cF}_2^{-1} \hat h_\alpha {\cF}_2,\quad  \alpha=1,2,3.
\end{equation*}
One has
\begin{equation*}
H=H_0-{V}_{1}-V_{2}-V_{3},
\end{equation*}
where $H_0$ is the multiplication operator by the function
$\sum_{\alpha =1}^3 \varepsilon _\alpha (k_\alpha)$
\begin{equation*}
(H_0f)(k_1,k_2,k_3)	 = \sum_{\alpha =1}^3 \varepsilon _\alpha
(k_\alpha)f(k_1,k_2,k_3), \quad f \in L_2(({\T}^3)^3).
\end{equation*}
The functions $\varepsilon_\alpha,\,\alpha=1,2,3 $ defined above are
of the form
\begin{equation}
\varepsilon _\alpha(p)=
l_\alpha\varepsilon(p),
\quad \varepsilon(p)=\sum_{i=1}^{3}(1-\cos p^{(i)}),\quad
p=(p^{(1)}, p^{(2)},p^{(3)})\in \R^3
\end{equation}
and $V_{\alpha},\alpha=1,2,3$  are integral operators of
convolution type
\begin{align*}
&(V_{\alpha}f)(k_1,k_2,k_3)
\\
& = \frac{\mu_\alpha}{(2\pi)^3}{\int\limits_{({\T}^3)^3} } \delta
(k_\alpha -k_\alpha ')\delta (k_\beta +k_\gamma -k_\beta
'-k_\gamma ')f(k'_1,k'_2,k'_3) dk'_1 dk_2'dk'_3,\\ &f \in
L_2(({\T}^3)^3),
\end{align*}
where $\delta (k)$ denotes the	Dirac delta-function.

For the two-particle Hamiltonians $h_\alpha,\alpha=1,2,3$ we have:
\begin{equation*}
h_\alpha =h_\alpha ^0-v_{\alpha},
 \end{equation*}
 where
\begin{equation*}
(h_\alpha ^0f)(k_\beta ,k_\gamma )=(\varepsilon _\beta (k_\beta)+
\varepsilon _\gamma (k_\gamma ))f(k_\beta ,k_\gamma )
\end{equation*}
and
\begin{equation}\label{interaction}
(v_{\alpha}f)(k_\beta ,k_\gamma )=
\frac{\mu_\alpha}{(2\pi)^3}{\int\limits_{({\T}^3)^2}}
 \delta (k_\beta
+k_\gamma -k_\beta '-k_\gamma ')f(k_\beta ',k_\gamma')dk_\beta
'dk_\gamma ',\quad f\in L_2(({\T}^3)^2).
\end{equation}
 \section{Decomposition of the energy operators into von Neumann direct integrals.
 Quasimomentum and coordinate systems}

 Given $m\in \N$, denote by $\hat U^m_s$, $s\in {\Z}^3$ the unitary operators on the Hilbert space
  $\ell_2(({\Z}^3)^m)$ defined as:
\begin{equation*}
(\hat U^m_sf)(n_1,n_2,..., n_m)=f(n_1+s,n_2+s,...,n_m+s),\quad
f\in \ell_2(({\Z}^3)^m).
\end{equation*}
We	easily see that
\begin{equation*}
  \hat U^m_{s+p}=\hat U^m_s\hat U^m_p,\quad s,p \in \Z^3,
  \end{equation*}
that is, $\hat U^m_s,s\in\Z^3 $ is a unitary  representation of
the abelian group $\Z^3.$

Via the Fourier transform $ \cF_m$ the unitary representation of
$\Z^3$ in  $\ell_2(({\Z}^3)^m)$ induces a representation of the
group $\Z^3$ in the Hilbert space $L_2(({\T}^3)^m)$	 by unitary
(multiplication) operators $U^m_s= \cF_m^{-1}\hat U^m_s\cF_m$,
$s\in \Z^3$ given by:
\begin{equation}\label{grup}
(U_s^mf)(k_1,k_2,...,k_m)= \exp \big (-i(s,k_1+k_2+...+k_m)\big
)f(k_1,k_2,...,k_m),
\end{equation}
\begin{equation*}
f\in L_2((\T^3)^m).
\end{equation*}

Decomposing the Hilbert space $ L_2((\T^3)^m)$	into the direct
integral
\begin{equation*}
L_2((\T^3)^m)= \int_{K\in {\T}^3} \oplus L_2(\F_K^m)d K,
\end{equation*}
where
\begin{equation*}
\F_K^m=\{(k_1,k_2,..., k_m){\ \in }({\T}^3)^m:k_1+k_2+...+k_m = K
(mod \, (2\pi\Z^1)^3)\},\quad K\in {\T}^3,
\end{equation*}
we obtain a corresponding decomposition of the unitary
representation $U_s^m$,
 $s \in \Z^3$ into the	direct integral
\begin{equation*}
U_s^m= \int_{K\in {\T}^3} \oplus U_s(K)d K,
\end{equation*}
where
\begin{equation*}
U_s(K)=\exp(-i(s,K))I \quad \text{on}\quad L_2(\F_K^m)
\end{equation*}
and $I=I_{L_2(\F_K^m)}$ denotes the identity operator on the
Hilbert space $ L_2(\F_K^m)$.

The above Hamiltonians $\widehat H$ and $\hat h_\alpha,\,
\alpha=1,2,3$ obviously commute with the groups of translations
$\hat U^3_s$  and $\hat U^2_s,\, s\in \Z^3$, respectively, that
is,
\begin{equation*}
\hat U^3_s\widehat H=\widehat H\hat U^3_s, \quad s\in \Z^3
\end{equation*}
and
\begin{equation*}
\hat U^2_s\hat h_\alpha=\hat h_\alpha\hat  U^2_s, \quad s\in \Z^3,
\quad \alpha=1,2,3.
\end{equation*}
Correspondingly, the Hamiltonians $ H$ and $ h_\alpha,\,
\alpha=1,2,3$ (in the momentum representation) commute with the
groups	$ U^m_s$, $s\in \Z^3$ given by \eqref{grup} for $m=3$ and
$m=2,$ respectively.

  Hence, the operators
  $H$ and $h_\alpha,\,\,
\alpha=1,2,3,$ can be decomposed into the direct integrals
\begin{equation*}
H=\int\limits_ {K \in {\T}^3}\oplus \widetilde H(K)dK \quad
\mbox{and}\quad h_\alpha= \int\limits_{k \in {\T}^3}\oplus\tilde
h_\alpha(k)d k, \quad \alpha=1,2,3,
\end{equation*}
with respect to	 the decompositions
\begin{equation*}
L_2 (( {\T}^3)^3) = \int\limits_ {K \in {\T}^3} {\ \oplus } L_2 (
\F_K^3 ) dK \quad \text{and}\quad L_2 (( {\T}^3)^2) = \int\limits_
{k \in {\T}^3} {\ \oplus } L_2 ( \F_k^2 ) d k,
\end{equation*}
respectively.

For any permutation $\alpha\beta\gamma$ of $123$
 we set:
\begin{equation}\label{mass}
l_{\beta\gamma}\equiv\frac{l_\beta}{l_\beta+l_\gamma},\quad
M\equiv\sum_{\alpha=1}^3\frac{1}{l_\alpha},\quad m_{\alpha}\equiv
\frac{1}{l_\alpha M},
\end{equation}
where the quantity $l_\alpha$ entered in Hypothesis 2.1.

Given a cyclic permutation ${\alpha}{\beta}{\gamma}$  of $123$ we
introduce the mappings
\begin{equation*}
\pi^{(3)}_\alpha:(\T^3)^3\to (\T^3)^2,\quad
\pi^{(3)}_\alpha((k_\alpha, k_\beta, k_\gamma))=(q_\alpha,
p_\alpha)
\end{equation*}
and
\begin{equation*}
\pi^{(2)}_\alpha:(\T^3)^2\to \T^3,\quad \pi^{(2)}_\alpha((k_\beta,
k_\gamma))=q_\alpha,
\end{equation*}
 where
\begin{align*}
&q_\alpha= l_{\beta\gamma} k_\beta-l_{\gamma\beta} k_\gamma \in
\T^3(mod \,( 2\pi \Z^1)^3)
 \quad
\text{ and } \\
& p_\alpha= m_\alpha (k_\beta+k_\gamma)-(m_\beta+m_\gamma)k_\alpha
\in \T^3(mod\, ( 2\pi \Z^1)^3).
\end{align*}

Denote by $\pi^{(3)}_{K}$ , $K\in \T^3$ resp. $\pi^{(2)}_k$, $k\in
\T^3$ the restriction of $\pi^{(3)}_\alpha$ resp.
$\pi^{(2)}_\alpha$ onto $\F_K^3\subset (\T^3)^3$ resp.
$\F_k^2\subset (\T^3)^2$, that is,
\begin{equation}\label{project}
\pi^{(3)}_{K}=\pi^{(3)}_\alpha\vert_{\F_K^3}\quad \text{and}\quad
\pi^{(2)}_{k}= \pi^{(2)}_\alpha\vert_{\F_k^2}.
\end{equation}
At this point it is useful to remark that
$$
\F^3_{K}=\{(k_\alpha,k_\beta ,k_\gamma )\in
({\bbT}^3)^2,k_\alpha+k_\beta +k_\gamma =K(mod\,( 2\pi \Z^1)^3)
\},\quad K \in {\bbT}^3
$$
and
$$
\F^2_{k}=\{(k_\beta ,k_\gamma )\in ({\bbT}^3)^2,k_\beta +k_\gamma
=k(mod\, ( 2\pi \Z^1)^3) \},\quad k \in {\bbT}^3
$$
are six and three-dimensional manifolds isomorphic to
${({\bbT}^3)^2}$ and ${\bbT}^3,$ respectively.

\begin{lemma}
The mappings $\pi^{(3)}_{K}$ , $K\in \T^3$ and $\pi^{(2)}_{k}$,
$k\in \T^3$ are bijective from $\F_K^3\subset (\T^3)^3$	 and
$\F_k^2\subset (\T^3)^2$ onto $(\T^3)^2$ and $\T^3$ with the
inverse mappings given by
\begin{equation*}
(\pi^{(3)}_{K})^{-1}(q_\alpha,p_\alpha)= (m_{\alpha} K-p_\alpha,
m_{\beta} K+l_{\gamma\beta}p_\alpha+ q_\alpha, m_{\gamma} K+
l_{\beta\gamma}p_\alpha- q_\alpha)
\end{equation*}
and
\begin{equation*}
(\pi^{(2)}_{k})^{-1}(q_\alpha)=(l_{\gamma\beta}k+
q_\alpha,l_{\beta\gamma}k-q_\alpha)\in (\T^3)^3.
\end{equation*}
\end{lemma}
\begin{proof} We obviously have that
\begin{equation*}
(m_{\alpha} K-p_\alpha)+ (m_{\beta} K+l_{\gamma\beta}p_\alpha+
q_\alpha)+ ( m_{\gamma} K+ l_{\beta\gamma}p_\alpha- q_\alpha)=K
\end{equation*}
and
\begin{equation*}
(l_{\gamma\beta}k+ q_\alpha)+(l_{\beta\gamma}k-q_\alpha)=k.
\end{equation*}
Therefore, the images of the mappings $(\pi^{(3)}_{K})^{-1}$ and
$(\pi^{(2)}_{k})^{-1}$ are the subsets of $\F^3_K$ and $\F^2_k,$
respectively.

Conversely, given
\begin{equation*}
(k_\alpha, k_\beta, k_\gamma)\in {\F}^3_K\subset (\T^3)^3 \quad
\text{and}\quad
 (k_\beta, k_\gamma)\in \F^2_k\subset (\T^3)^2
\end{equation*}
one computes that
\begin{align*}
(\pi^{(3)}_{K})^{-1}(q_\alpha,p_\alpha)= (k_\alpha, k_\beta,
k_\gamma)
 \quad \text{ and } \quad
(\pi^{(2)}_{k})^{-1}(q_\alpha)=(k_\beta, k_\gamma),
\end{align*}
where
\begin{align*}
& q_\alpha\equiv l_{\beta\gamma} k_\beta-l_{\gamma\beta} k_\gamma
\in (\T^3)(mod\, ( 2\pi \Z^1)^3) \quad \text{ and } \\&
p_\alpha\equiv m_\alpha
(k_\beta+k_\gamma)-(m_\beta+m_\gamma)k_\alpha\in (\T^3)(mod\, (
2\pi \Z^1)^3) .
\end{align*}
\end{proof}

The fiber operator $\widetilde H(K),$ $K \in {\bbT}^3$ is
unitarily equivalent to the operator  $$
H(K)=H_0(K)-V_1-V_2- V_3.
 $$
In the coordinates $(q_\alpha,p_\alpha)$
 the operators $H_0(K)$
and $V_\alpha$ are defined on the Hilbert space $L_2 ((
{\T}^3)^2)$ by
$$(H_0(K)f)(q_\alpha,p_\alpha)=E_{\alpha\beta }
(K;q_\alpha,p_\alpha)f(q_\alpha,p_\alpha),\quad f\in L_2 ((
{\T}^3)^2), $$

 \be\label{potential} (V_\alpha
f)(q_\alpha,p_\alpha)=\frac{\mu_\alpha}{(2\pi)^3}\int\limits_{{\bbT}^3}
f(q_\alpha',p_\alpha)dq'_\alpha,\quad f\in L_2 (({\T}^3)^2), \ee
 where
$$ E_{\alpha\beta}(K;q_\alpha, p_\alpha)= \varepsilon _\alpha
(m_{\alpha}K-p_\alpha)+\varepsilon _\beta (m_{\beta} K
+l_{\gamma\beta} p_\alpha+q_\alpha) +\varepsilon _\gamma
(m_{\gamma} K+l_{\beta\gamma} p_\alpha-q_\alpha).
$$
The fiber operator $\tilde h_\alpha(k),$ $k \in {\bbT}^3,$
 $\alpha=1,2,3$
is unitarily equivalent to the operator
\begin{equation}\label{two} h_\alpha (k) =h_\alpha^{0}
(k)-v_{\alpha},
\end{equation} where
$$(h_\alpha^{0}
(k)f)(q_\alpha)=E_{k}^{(\alpha)}(q_\alpha)f(q_\alpha),f \in
L_2({\bbT}^3),
$$
\begin{equation} (v_\alpha
f)(q_\alpha)=\frac{\mu_\alpha}{(2\pi)^3}\int\limits_{{\bbT}^3}
f(q')d q', \quad f \in L_2({\bbT}^3)
\end{equation}
and
\begin{equation}\label{E-alpha}
E_{k}^{(\alpha)}(q_\alpha)= \varepsilon _\beta
(l_{\gamma\beta}k+q_{\alpha}) +\varepsilon _\gamma
(l_{\beta\gamma}k-q_{\alpha}).
\end{equation}

Let
$$
U_K :L_2({ {\F}^3_K}) \longrightarrow L_2(({\T}^3)^2) ,
 \,\, U_K f=f \circ (\pi^{(3)}_{K})^{-1},
\,\, K \in
{\bbT}^3,
$$
and
$$
u_{k}:L_2(\F^2_{k}) \rightarrow L_2({\bbT}^3),	\,\,
u_{k} g=g \circ(\pi^{(2)}_{k})^{-1},\,k\in {\bbT}^3,
$$
where $\pi^{(3)}_{K}$ and $\pi^{(2)}_{k}$ are defined by
\eqref{project}. Then $U_K$ and $u_k$ are unitary operators and
\begin{equation*}
H(K) = U_K \widetilde H(K)U_K^{-1},\quad h_\alpha (k) =u_{k}
\tilde h_\alpha (k)u_{k}^{-1},\quad \alpha=1,2,3.
\end{equation*}

All of our further calculations will be carried out in the
"momentum representation" in a system of coordinates connected
with the fixed center of inertia of the system of three particles.
 We order 1,2,3 by the
conditions $1 \prec 2,$ $2 \prec 3$ and $3 \prec 1.$
 Sometimes instead of the
coordinates $(q_{\alpha},p_{\alpha})$ (if it does not lead to any
confusion we will write $(q,p)$ instead of $(q_\alpha,p_\alpha)$)
it is convenient to choose some pair of the three variables
$p_{\alpha}.$ The connection between the various coordinates is
given by the relations
\begin{equation}\label{coordinate}
p_1+p_2+p_3=0,\\
\pm q_\alpha=l_{\gamma\beta}p_\alpha + p_\beta, \quad
l_{\gamma\beta}=\frac{l_\gamma}{l_\beta+l_\gamma},\quad
(\alpha\neq\beta, \beta\neq \gamma,\gamma\neq\alpha),
\end{equation} where the plus sign corresponds to the case $\beta
\prec \alpha,$ the minus sign corresponds to the case $\alpha
\prec \beta.$ Expressions for the variables $q_{\alpha}$ in terms
of $p_{\alpha}$ and $p_{\beta}$ can be written in the form
$q_{\alpha}=d_{\alpha\beta}p_{\alpha}+e_{\alpha\beta}p_{\beta}$
and explicit formulas for the coefficients $d_{\alpha\beta}$ and
$e_{\alpha\beta}$ are obtained by combining the latter equation
with \eqref{coordinate}.

\section{Statement of the main results}

For each $K\in\T^3$ the minimum and the maximum taken over $(q,p)$
of the function $E_{\alpha \beta}(K;q,p) $ are independent of ${\
\alpha },{\ \beta } =1,2,3.$ We set:
\begin{align*}
E _{\min }(K)\equiv\min_{q,p}E_{\alpha \beta }(K,q,p),\quad E_{max
}(K)\equiv\max_{q,p}E_{\alpha \beta }(K,q,p).
\end{align*}

\begin{definition} \label{def}The operator $h_{\alpha}(0)$ is said to
have a zero energy resonance if the equation
$$
\frac{(2\pi)^{-3}\mu_\alpha}{l_\beta+l_\gamma}\int\limits_{\T^3}\,
(\varepsilon(q'))^{-1} {\varphi(q')dq'}=\varphi(q)
$$ has a nonzero solution $\varphi$ in the Banach space $C(\T^3).$
Without loss of generality we can always normalize
$\varphi $ so that $\varphi(0)=1.$
\end{definition}

Let the operator $h_{\alpha}(0)$ have a zero energy resonance.
Then the function
$$
\psi(q)=(\varepsilon(q))^{-1}
$$
is a solution (up to a constant factor) of the Schr\"{o}dinger
equation $h_\alpha(0)f= 0$ and $\psi$ belongs to $L_1({\bbT}^3)$.

Set \be
\mu_\alpha^0=(l_\beta+l_\gamma)(2\pi)^3(\int\limits_{{\bbT}^3}(\varepsilon(q))^{-1}
 {d q})^{-1}. \ee
\begin{hypothesis}\label{hypoth}
We assume that
$\mu_\alpha=\mu^{0}_\alpha,\, \mu_\beta=\mu^{0}_\beta $ and
$\mu_\gamma \leq \mu^{0}_\gamma$.
\end{hypothesis}

The main results of the paper are given in the following theorems.
\begin{theorem}\label{esss} For the essential spectrum
${\sigma}_{ess} ( H(K))$ of $H(K)$ the following equality
$$
\sigma_{ess}(H (K))=\cup^3_{\alpha=1}\cup _{p\in {\bbT}^3}\{\sigma
_d(h_{\alpha}
 ((m_\beta+m_\gamma)K+p))+\varepsilon _\alpha
(m_{\alpha} K-p)\} \cup [E_{\min}(K),E_{\max}(K)],
$$
holds,	where $\sigma _d (h_{\alpha}(k))$ is the discrete spectrum
of the operator $h_{\alpha}(k),k\in \T^3$.
\end{theorem}
Denote by $\tau_{s}(K)$ the bottom of the spectrum of the
self-adjoint bounded operator $H(K)$, that is,
$$\tau_{s}(K)= \inf_{||f||=1}(H(K)f,f).$$

We set:
\begin{equation}\label{spkanal}
\tau^{\gamma}_\text{s}(K)\equiv\inf_{||f||=1}[(H_0(K)f,f)-(V_\gamma
f,f)],\gamma=1,2,3.\end{equation}

As in the introduction, let $N(K,z)$ denote the number of
eigenvalues of the operator $H(K),\,K\in {\bbT}^3$ below $z \leq
\tau_{ess}(K),$ where $\tau_{ess}(K)\equiv\inf\sigma_{ess}(H (K))$
is the bottom of the essential spectrum of $H(K).$

\begin{theorem} \label{baho} Assume Hypothesis \ref{hypoth}.
 Then  for all $K\in \T^3$
 the inequality
$$\tau^{\alpha}_\text{s}(K)-\mu^0_\beta-\mu_\gamma\leq \tau_{s}(K)$$
holds.
\end{theorem}
Theorem \ref{baho} yields the following
\begin{corollary} Assume Hypothesis \ref{hypoth}.
All eigenvalues of the operator $H(K),K\in \T^3$ below the bottom
of $\tau_\text{ess}(K)$ belong to the interval
$[\tau^{\alpha}_\text{s}(K)-\mu^0_\beta-\mu_\gamma,\tau_\text{ess}(K)).$
\end{corollary}

\begin{theorem}\label{finite} Assume Hypothesis \ref{hypoth}. Then for all $ K \in
U^{0}_\delta(0),$ $\delta>0$ sufficiently small,
 the operator $H(K)$ has a finite number of eigenvalues below the
 bottom of the
essential spectrum of $H(K)$.
\end{theorem}
\begin{theorem}\label{asym.zK} Assume Hypothesis \ref{hypoth}.
Then the operator $H(0)$ has infinitely many eigenvalues below the
bottom of the essential spectrum and the functions $N(0,z)$ and
$N(K,0)$ obey the relations
\begin{equation} \label{asym.K}
\lim\limits_{z \to -0}\frac{N(0,z)}{|\log |z||}=\lim\limits_{|K|
\to 0}\frac{N(K,0)} {2|\log |K||}={\cU}_0  \,(0<{\cU}_0 <\infty).
\end{equation}
\end{theorem}
\begin{remark} The constant ${\cU}_0$ does not depend on the
pair potentials $\mu_{\alpha},\,\alpha=1,2,3$ and is given as a positive function
depending only on the ratios $\frac {l_\beta}{l_\alpha},\,\alpha\neq \beta,\,\alpha,\beta=1,2,3$ between
the masses.
\end{remark}

\section{ Spectral properties of the
two-particle operator $h_\alpha(k)$}

In this section we study the spectral properties of the
two-particle discrete Schr\"{o}dinger operator $h_\alpha(k),$
$k\in {\T^3}.$

We consider the family of the self-adjoint operators $h_\alpha(k),
\,\, k\in {\bbT}^3$	 on the Hilbert space $L_2 ({\bbT}^3)$
\begin{equation}
 h_\alpha(k)=h^0_\alpha (k)-\mu_\alpha v.
\end{equation}

The nonperturbed operator $h^0_\alpha (k)$ on $L_2({\bbT}^3)$ is
multiplication operator by the function $ E^{(\alpha)}_k(p)$
$$
(h^0_\alpha (k)f)(p)=E^{(\alpha)}_k (p) f(p),\quad f\in L_2
({\bbT}^3),$$ where $ E^{(\alpha)}_k (p)$ is defined in
\eqref{E-alpha}.
 The perturbation $v$ is an integral
operator of rank one
$$(v f)(p)=(2\pi)^{-3}
\int\limits_{{\bbT}^3} f(q)dq,\quad f\in L_2 ({\bbT}^3).$$

Therefore by the Weyl theorem the continuous spectrum
   $\sigma_{\text{cont}}(h_\alpha(k))$ of the
operator $h_\alpha(k),\,k \in \T^3$ coincides with the spectrum $
{\sigma}( h^0_\alpha (k) ) $ of $h^0_\alpha(k).$ More
specifically,
$$
 \sigma_{\text{cont}}(h_\alpha(k))= [E^{(\alpha)}_{\min}(k)
,\,E^{(\alpha)}_{\max}(k)],
$$
 where
\be\label{min} E^{(\alpha)}_{\min }(k)\equiv\min_{p\in
\T^3}E^{(\alpha)}_k(p),\quad
E^{(\alpha)}_{\max}(k)\equiv\max_{p\in {\bbT}^3}
E^{(\alpha)}_{k}(p). \ee


\begin{lemma}
There exist an odd and analytic function $p_\alpha:\T^3\rightarrow
\T^3 $ such
  that for any $k \in (\pi,\pi]^3$	the point $p_\alpha(k) $ is a unique
  non degenerate minimum  of the function
$E^{(\alpha)}_k(p)$ and
\begin{equation}\label{Okatta}
p_\alpha(k)=O(|k|^3)\,\,\mbox{as}\,\,k \to 0.
\end{equation}
\end{lemma}

\begin{proof}
The function $E^{(\alpha)}_k(p)$ can be rewritten in the form
\begin{equation}\label{echiladigan}
 E^{(\alpha)}_k(p)=3(l_\beta+l_\gamma)-\sum_{j=1}^3
 (a_\alpha(k^{(j)})\cos p^{(j)}+b_\alpha(k^{(j)})\sin p^{(j)}),
\end{equation}
where the coefficients $a_\alpha(k^{(j)})$ and $b_\alpha(k^{(j)})$
are given by
\begin{equation}\label{formul}
a_\alpha(k^{(j)})=l_\beta\cos(l_{\gamma\beta}k^{(j)})+
l_\gamma\cos(l_{\beta\gamma}k^{(j)}),
b_\alpha(k^{(j)})=l_\beta\sin(l_{\gamma\beta}k^{(j)})-
l_\gamma\sin(l_{\beta\gamma}k^{(j)}).
\end{equation} The equality \eqref{echiladigan} implies the following
representation for $E^{(\alpha)}_k(p)$
\begin{equation}\label{represent}
E^{(\alpha)}_k(p)=3(l_\beta+l_\gamma)-\sum_{j=1}^3r_\alpha(k^{(j)})\cos
(p^{(j)}- p_\alpha(k^{(j)})),
\end{equation}
where
$$
r_\alpha(k^{(j)})=\sqrt{a^2_\alpha(k^{(j)})+b^2_\alpha(k^{(j)})},\quad
p_\alpha(k^{(j)})=\arcsin\frac{b_\alpha(k^{(j)})}{r_\alpha(k^{(j)})},\quad
k^{(j)}\in (-\pi,\pi].$$
 Taking into account \eqref{formul}, we have that the vector-function
$$p_\alpha:\T^3\rightarrow \T^3,\quad p_\alpha=p_\alpha(k^{(1)},k^{(2)},k^{(3)})=
(p_\alpha(k^{(1)}),p_\alpha(k^{(2)}),p_\alpha(k^{(3)}))\in\T^3$$
is odd regular and it is the minimum point
 of $E^{(\alpha)}_k(p).$ One has, as easily seen from the definition  $$
p_\alpha(k)=O(|k|^3)\,\,as\,\,k \to 0.
$$
\end{proof}
 Let {\bf C} be the complex plane.
 For any $k \in \T^3$ and $z{\in } {\bf C} { \setminus } { \sigma
 }_\text{cont}(h_\alpha(k))$ we define
a	function (the Fredholm's determinant
 associated with the operator $h_\alpha(k)$)
\begin{equation}
{\Delta}_\alpha( k, z) = 1-\mu_\alpha(2\pi)^{-3}
\int\limits_{{\T}^3}(E^{(\alpha)}_k (q) -z )^{-1}d q .
\end{equation}

Note that the function ${\Delta}_\alpha( k, z)$ is real-analytic
in $(-\pi,\pi]^3\times ({\bf C} { \setminus } { \sigma
 }_\text{cont}(h_\alpha(k)))$

The following lemma is a simple consequence of the
Birman-Schwinger principle and Fredholm's theorem.

\begin{lemma}\label{nollar}
Let	 $k { \in } \T^3.$	The point $z { \in } {\bf C} { \setminus }
 {\sigma}_\text{cont}(h_\alpha(k))$	   is an
eigenvalue of the operator $h_\alpha(k) $ if and only if
$$
  {\Delta_\alpha}(k, z) = 0.
 $$
 $\Box$
\end{lemma}

\begin{lemma}\label{resonance}	The following statments are equivalent:\\
(i) the operator $h(0)$ has
a zero energy resonance;\\
(ii)  $\Delta_\alpha(0,0)=0;$\\
(iii) $\mu_\alpha= \mu_\alpha^{0}.$
\end{lemma}
  \begin{proof} Let the operator $h_\alpha(0)$ has
a zero energy  resonance for some $\mu_\alpha>0$. Then by
 \eqref{def} the equation
\begin{equation}
 \varphi (p)=\mu_\alpha(l_\beta+l_\gamma)^{-1}(2\pi)^{-3}\int\limits_{{\bbT}^3}(\varepsilon(q))^{-1}
 \varphi (q)dq
\end{equation}
 has a simple solution in $C({\T^{3}}) $ and
 the solution
$\varphi (q)$ is equal to $1$ (up to a constant factor).
 Therefore we see that
 $$
1=\mu_\alpha(l_\beta+l_\gamma)^{-1}(2\pi)^{-3}\int\limits_{{\bbT}^3}(\varepsilon(q))^{-1}
 {d q}
 $$
and hence $$ \Delta_\alpha(0,0)=
1-\mu_\alpha(l_\beta+l_\gamma)^{-1}(2\pi)^{-3}\int\limits_{{\bbT}^3}(\varepsilon(q))^{-1}
 {d q}=0
$$ and so
$\mu_\alpha=\mu_\alpha^{0}$.

 Let for some $\mu_\alpha>0$ the equality $$\Delta_\alpha(0,0)=
1-\mu_\alpha(l_\beta+l_\gamma)^{-1}(2\pi)^{-3}\int\limits_{{\bbT}^3}(\varepsilon(q))^{-1}
 {d q}=0 $$	 holds and consequently $\mu_\alpha=\mu_\alpha^{0}$.
Then  only the function $\varphi (q)\equiv
 constant \in  C({\T}^{3})$ is a solution of the equation
 $$
 \varphi (p)=\mu_\alpha(l_\beta+l_\gamma)^{-1}(2\pi)^{-3}\int\limits_{{\T}^3}
  (\varepsilon(q))^{-1}{\varphi (q)dq}
,$$ that is, the operator $h_\alpha(0)$ has a zero energy
resonance.
\end{proof}

\begin{theorem}\label{mavjud} Let the
operator $h_\alpha(0)$ have a  zero energy resonance. Then	for
all $k \in {\T}^3,\, k \neq 0$ the operator $h_\alpha(k)$ has a
unique simple eigenvalue $z_\alpha(k)$	below the bottom of the
continuous spectrum of $h_\alpha(k).$ Moreover $z_\alpha(k)$ is
even on ${\T}^3$ and $z_\alpha(k)>0$ for $k\neq 0.$
\end{theorem}
\begin{proof}
 By Lemma \ref{resonance}

 $$\Delta_\alpha(0,0)=
1-\mu_\alpha^0(l_\beta+l_\gamma)^{-1}(2\pi)^{-3}\int\limits_{{\bbT}^3}(\varepsilon(q))^{-1}
 {d q}=0
$$
and hence it is easy to see that for any $z<0$ the inequality
$\Delta_\alpha(0,z)>0$ holds. By Lemma \ref{nollar} the operator
$h_\alpha(0)$ has no negative eigenvalues. Since $p=p_\alpha(k)$
is the non degenerate minimum of the function $E^{(\alpha)}_k(p)$
we define $\Delta_\alpha(k,E^{(\alpha)}_{\min}(k))$ as
$$\Delta_\alpha(k,E^{(\alpha)}_{\min}(k))
=1-\mu_\alpha^{0}(2\pi)^{-3}\int\limits_{{\T}^3}(E^{(\alpha)}_k(q)-E^{(\alpha)}_{\min}(k))^{-1}{dq}.
$$
By	dominated convergence theorem we have
\begin{equation}
\lim_{z\to
{\tiny E^{(\alpha)}_{\min}(k)}}\Delta_\alpha(k,z)=\Delta_\alpha(k,E^{(\alpha)}_{\min}(k)).
\end{equation}
For all $k\neq 0,
 q\neq 0$  the inequality
 $$E^{(\alpha)}_k(q+p_{\alpha}(k))-E^{(\alpha)}_{\min}(k)<E_0(q)$$ holds and
 hence we obtain the following inequality
\begin{equation}\label{xos qiymat}
\Delta_\alpha(k,E^{(\alpha)}_{\min}(k))< \Delta_\alpha(0,0)=0,
k \neq 0.
\end{equation}
For each $k \in {\T}^3$ the function $\Delta_\alpha(k,\cdot)$ is
monotone decreasing	 on $(-\infty,E^{(\alpha)}_{\min}(k)]$ and
$\Delta_\alpha(k,z)\to 1$ as $z \to -\infty.$  Then by virtue of
\eqref{xos qiymat} there is a number $z_\alpha(k)\in
(-\infty,E^{(\alpha)}_{\min}(k))$ such that
$\Delta_\alpha(k,z_\alpha(k))=0.$
 By Lemma \eqref{nollar} for any nonzero
$k\in\T^3$ the operator $h_\alpha(k)$ has an eigenvalue below
$E^{(\alpha)}_{\min}(k)$. For any $k \in {\T}^3$ and
$z \in (-\infty,E^{(\alpha)}_{\min}(k))$ the equality
$\Delta_\alpha(-k,z)=\Delta_\alpha(k,z)$ holds and hence
$z_\alpha(k)$ is even.

Let us	prove the positivity of the eigenvalue $z_\alpha(k),\,k\neq 0$.
First we verify, for all $ k \in {\T}^3,k\neq 0,$  the inequality
\begin{equation}\label{musbat}
  \Delta_\alpha
(k,0)>0.
\end{equation}
Applying the definition of $\mu_\alpha^0$ by (4.1) and making a
change of variables $p=l_{\gamma\beta}k+q$ in \eqref{musbat} we have
\begin{equation}\label{d(k,0)} \Delta_\alpha(k,0)= \mu_\alpha^0 (2\pi)^{-3}\int\limits_{{\T}^3}
\frac{ \varepsilon_\gamma(k-p)- \varepsilon_\gamma(p)} {E_0(p)
(\varepsilon_\beta (p)+\varepsilon_\gamma(k-p))} d p.
\end{equation} Making a change of variables $q=\frac{k}{2}-p$ in
\eqref{d(k,0)} and using the equality
$\Delta_\alpha(k,0)=\Delta_\alpha(-k,0)$ it is easy to show that
$$
\Delta_\alpha(k,0)= \frac{\Delta_\alpha (k,0)+\Delta_\alpha
(-k,0)} {2}=
$$
$$
=\frac{\mu_\alpha^{0}}{2}(2\pi)^{-3}\int\limits_{{\T}^3}
 (\varepsilon_\gamma(\frac{k}{2}+p)- \varepsilon_\gamma(\frac{k}{2}-p))
 (\varepsilon_\beta(\frac{k}{2}+p)-
\varepsilon_\beta(\frac{k}{2}-p)) F(k,p)dp,
$$
where
$$
F(k,p)=\frac{E_0(\frac{k}{2}+p)+ E_0(\frac{k}{2}-p)}
{E_0(\frac{k}{2}+p) E_0(\frac{k}{2}-p) (\varepsilon_\beta
(\frac{k}{2}+p)+\varepsilon_\gamma(\frac{k}{2}-p))
(\varepsilon_\beta
(\frac{k}{2}-p)+\varepsilon_\gamma(\frac{k}{2}+p))}>0.
$$
A simple computation shows that
  $$
 (\varepsilon_\gamma(\frac{k}{2}+p)- \varepsilon_\gamma(\frac{k}{2}-p))
(\varepsilon_\beta(\frac{k}{2}+p)-
\varepsilon_\beta(\frac{k}{2}-p))=4l_\beta l_\gamma \left(
\sum_{i=1}^{3}cos\frac{k^{(i)}}{2}cos p^{(i)} \right)^2 \geq 0.
$$
Thus  the inequality \eqref{musbat} is proven.

For any $k \in {\T}^3$ the function $\Delta_\alpha(k,\cdot)$ is
monotone decreasing and the inequalities
$$\Delta_\alpha(k,0)>\Delta_\alpha(k,z_\alpha(k))=0
>\Delta_\alpha(k,E^{(\alpha)}_{\min}(k)),\text\quad k\neq 0$$ hold. Therefore
 the eigenvalue $z_\alpha(k)$ of the operator  $h_\alpha(k)$
belongs to	interval $(0,E^{(\alpha)}_{\min}(k)).$
\end{proof}

 The
following decomposition is important for the proof of the
asymptotics	 \eqref{asym.zK}.

 \begin{lemma}\label{razlojeniya}Let $\mu_\alpha=\mu_\alpha^0,\alpha=1,2,3.$
 Then for any $k\in U_{\delta}(0),\delta>0$ sufficiently small, and
 $z\leq E^{(\alpha)}_{\min}(k)$	 the following decomposition holds:
\begin{align*}
&\Delta_\alpha(k,z)= \frac{\mu_\alpha^0 }{\sqrt{2}\pi (l_\beta
+l_\gamma)^{\frac{3}{2}}} \left [E^{(\alpha)}_{\min}(k)-z \right
]^{\frac{1}{2}}\\
&+ \Delta^{(20)}_\alpha(E^{(\alpha)}_{\min}(k)-z
)+\Delta^{(02)}_\alpha(k,z)\quad\text{as}\quad z \to
E^{(\alpha)}_{\min}(k),
\end{align*}
where
$\Delta^{(20)}_\alpha(E^{(\alpha)}_{\min}(k)-z)=O(E^{(\alpha)}_{\min}(k)-z)$
and $\Delta^{(02)}_\alpha(k,z)=O(|k|^2)\quad \text{as}\quad
k\rightarrow 0.$
\end{lemma}

\begin{proof}
 Let
 \begin{equation}
 E_\alpha(k,p)=E^{(\alpha)}_k
(p+p_\alpha(k))-E^{(\alpha)}_{\min}(k),
\end{equation}
where $p_\alpha(k)\in\T^3$ is the minimum point of the function
$E^{(\alpha)}_k (p),$ that is,
$E^{(\alpha)}_{\min}(k)=E^{(\alpha)}_k (p_\alpha(k))$. Then using
\eqref{represent} we conclude
\begin{equation*}
E_\alpha(k,p)=\sum_{j=1}^3r_\alpha(k^{(j)})(1-\cos p^{(j)}).
\end{equation*}
We define the function $\tilde\Delta_\alpha(k,w)$ on
$(-\pi,\pi]^3\times{\R}^1$ by $
\tilde\Delta_\alpha(k,w)=\Delta_\alpha(k,E^{(\alpha)}_{\min}(k)-w^2).$
The function $\tilde\Delta_\alpha(k,w)$ represented as
\begin{align*}
&\tilde\Delta_\alpha(k,w)=1-\mu_\alpha(2\pi)^{-3}\int\limits_{{\T}^3}
\frac{dp} {E_\alpha(k,p)+w^2}\\&=1-\mu_\alpha
(2\pi)^{-3}\int_{\T^3}\frac{dp}{\sum_{j=1}^3r_\alpha(k^{(j)})(1-\cos
p^{(j)})+w^2}\end{align*} and so it is real-analytic in
$(-\pi,\pi]^3\times{\R}^1$ and even in $k\in(-\pi,\pi]^3.$
Therefore
 \be
 \tilde\Delta_\alpha(k,w)=\tilde\Delta_\alpha(0,w)+\tilde\Delta^{(20)}_\alpha(k,w),
\ee
 where $\tilde\Delta^{(20)}_\alpha(k,w)=O(|k|^2)$
 uniformly in  $w \in {\R^1}$ as $k\to 0.$
Taylor series expansion gives
 $$
\tilde\Delta_\alpha(0,w)=\tilde\triangle^{(01)}_\alpha(0,0)w+
\tilde\triangle^{(02)}_\alpha(0,w)w^2 $$ where
$\tilde\triangle^{(02)}_\alpha(0,w)=O(1)\quad\text{as}\quad w\to
0.$ Simple computation shows that
\begin{equation}\label{partial}\frac{d\tilde\Delta_\alpha(0,0)}{d
w}=\tilde\triangle^{(01)}_\alpha(0,0)= \frac{\mu_{\alpha}^0}{\sqrt{2}
\pi(l_\beta +l_\gamma)^ {\frac{3}{2}}}\neq 0.\end{equation}
\end{proof}
\begin{corollary}\label{corol}
The function $z_\alpha(k)=E^{(\alpha)}_{\min}(k)-w^2_\alpha(k)$ is
real-analytic in $U_\delta(0),$ where $w_\alpha(k)$ is a unique simple
solution of the equation $\tilde\Delta_\alpha(k,w)=0$ and
$w_\alpha(k)=O(|k|^2)$ \text{as}\quad $k\to 0$.
\end{corollary}
\begin{proof}
 Since
 $\tilde\Delta_\alpha(0,0)=0$ and the inequality \eqref{partial}
holds the equation $\tilde\Delta_\alpha(k,w)=0$ has a unique
simple solution $w_\alpha(k),\,k\in U_\delta(0)$ and it is real-analytic
in $U_\delta(0).$ Taking into account that the function
$\tilde\Delta_\alpha(k,w)$ is even in $k\in U_\delta(0),\delta>0$
and $w_\alpha(0)=0$ we have that $w_\alpha(k)=O(|k|^2).$ Therefore
the function $z_\alpha(k)=E^{(\alpha)}_{\min}(k)-w^2_\alpha(k)$ is
real-analytic in $U_\delta(0).$
\end{proof}
 \begin{lemma}\label{tasvir} Let
 $\mu_\alpha=\mu_\alpha^0$ for some $\alpha=1,2,3$.
 Then for any $k \in U_{\delta}^0(0)$ there exists a number
 $\delta(k)>0$ such that, for all $z \in
 V_{\delta(k)}(z_\alpha(k))$, where $V_{\delta(k)}(z_\alpha(k))$
 is the	 $\delta(k)$-neighborhood of the point	$z_\alpha(k)$,
 the following representation holds
$$
\Delta_\alpha(k,z)=C_1(k)(z-z_\alpha(k))\hat\Delta_\alpha(k,z).
$$
Here $C_1(k)\neq 0$ and $\hat\Delta_\alpha(k,z)$ is continuous in
$V_{\delta(k)}(z_\alpha(k))$ and
$\hat\Delta_\alpha(k,z_\alpha(k))\neq 0.$
\end{lemma}

\begin{proof}
Since $z_\alpha(k)<E^{(\alpha)}_{\min}(k),\,k\neq0$ the function
$\Delta_{\alpha}(k,z)$ can be expanded as follows
 $$
\Delta_\alpha(k,z)=\sum_{n=1}^{\infty}C_n(k)(z-z(k))^n, \quad z\in
V_{\delta(k)}(z_{\alpha}(k)),$$
   where
$$ C_1(k)= \frac{\mu_\alpha^o}{ \sqrt{2}\pi(l_\beta
+l_\gamma)^{\frac{3}{2}}}
\frac{1}{2\sqrt{E_{\min}^{(\alpha)}(k)-z_\alpha(k)}}\neq 0,\quad
k\neq 0.
$$

Therefore $\hat\Delta_\alpha(k,z)$ is continuous in
$V_{\delta(k)}(z_{\alpha}(k))$.
 Since $z_\alpha(k),\, k\neq 0$ is an unique simple solution of the
equation $\Delta_{\alpha}(k,z)=0,\,z\leq E_{\min}^{(\alpha)}(k)$,
we have	 $\hat\Delta_\alpha(k,z_\alpha(k))\neq 0.$

\end{proof}

\section{Spectrum of the operator $ H(K)$}
The "channel operator" $H_{\alpha} ( K), K {\ \in } {\T}^3$ acts
in the Hilbert space $L_2 (({\T}^3 )^2) $ as
\begin{equation}
 H_{\alpha}(K)=H_0(K)- V_{\alpha}.
\end{equation}

The decomposition of the space $L_2 (( {\T}^3 ) ^2) $ into the
direct integral

 $$L_2 (( {\T}^3 ) ^2) = \int\limits_{p\in {\T}^3}
\oplus L_2({\T}^3) d p
$$

yields for the operator $H_{\alpha}(K)$
the decomposition into the direct integral
 $$ H_{\alpha}(K) =	 \int\limits_{p\in {\T}^3}
 \oplus H_{\alpha}(K,p) dp.$$
 The fiber operator $H_{\alpha}(K,p)$ acts in the
Hilbert space $L_2({\bbT}^3)$ and has the form
\begin{equation*}
 H_{\alpha}(K,p) =h_{\alpha}
((m_\beta+m_\gamma)K+p)+\varepsilon_{\alpha} (m_{\alpha}K-p)
I,
\end{equation*} where I is
identity operator and $h_{\alpha}(k)$ is the two-particle operator
defined
  by (5.1).
The representation of the operator $H_\alpha (K,p)$ implies the
equality
\begin{align*}\label{stucture}
 &\sigma (H_\alpha (K,p))
 = \sigma _d(h_{\alpha}((m_\beta+m_\gamma)K+p))\\
 &\cup \big[E_{\min
}^{(\alpha)} ((m_\beta+m_\gamma)K+p),E_{\max}^{(\alpha
)}((m_\beta+m_\gamma)K+p) \big]+\varepsilon _\alpha
(m_{\alpha}K-p),
\end{align*}
 where $\sigma _d(h_{\alpha}(k))$
 is the discrete spectrum of the operator
 $h_{\alpha}(k).$
The theorem (see, e.g.,[22]) on the spectrum of decomposable
operators and above obtained structure for the spectrum of
$H_\alpha (K,p)$ gives
 \begin{lemma}\label{spec} The
equality holds
\begin{align*}
&\sigma (H_\alpha (K))\\&=\cup _{p\in {\T}^3} \left \{ \sigma
_d(h_{\alpha}((m_\beta+m_\gamma)K+p)+ \varepsilon _\alpha
(m_{\alpha}K-p) \right\} \cup [E_{\min}(K),E_{\max}(K)].
\end{align*}
\end{lemma}

\begin{lemma}\label{inequality} Let $\mu_\alpha=\mu_\alpha^0$ for some
$\alpha=1,2,3.$ Then for any  $K\in U^0_\delta(0),\delta>0$
sufficiently small,
 the following
inequality
$$\tau^\alpha_s(K)<E_{\min} (K)$$ holds, where $\tau^\alpha_s(K)$
is defined in \eqref{spkanal}.
\end{lemma}
\begin{proof} By Theorem \ref{mavjud} for each $K\in {\T}^3$
and $p\in {\T}^3,$ $p\not= -(m_\beta+m_\gamma)K$ the operator
$h_{\alpha} ((m_\beta+m_\gamma)K+ p)$ has an unique simple
positive eigenvalue $z_\alpha ((m_\beta+m_\gamma)K+p)$ below the
bottom of $\sigma _{cont}(h_{\alpha} ((m_\beta+m_\gamma)K+p)).$
Therefore using Lemma \ref{spec} for the spectrum $\sigma
(H_{\alpha} (K))$ of the operator $H_{\alpha}(K)$
 we conclude that
\begin{align*}
&\tau^\alpha_s(K)=
 \inf\cup_{p\in {\T}^3} \left
 [ \varepsilon_{\alpha}(m_{\alpha} K-p)
+  \sigma (h_{\alpha} ((m_\beta+m_\gamma)K+
p))\right]\\
&= \min_{p\in {\T}^3} \left [ \varepsilon_{\alpha}(m_{\alpha}
K-p)+ z_\alpha ((m_\beta+m_\gamma)K+p)\right].
\end{align*}
By Theorem \ref{mavjud} for each $K\in U^0_\delta(0)$ and
$p\neq-(m_\beta+m_\gamma)K$ the inequality
$$\varepsilon_{\alpha}(m_{\alpha} K-p)+ z_\alpha
((m_\beta+m_\gamma)K+p)< E_{\min }^{(\alpha)}
((m_\beta+m_\gamma)K+p)+\varepsilon_{\alpha}(m_{\alpha} K-p)$$
holds.

On the other hand, by computing partial derivatives, it is easy to
see that for any $K\in U^0_\delta(0)$ the point
$p=-(m_\beta+m_\gamma)K$ can not be a minimum point for the
function $E_{\min }^{(\alpha)}
((m_\beta+m_\gamma)K+p)+\varepsilon_{\alpha}(m_{\alpha} K-p).$
 Therefore $\tau^\alpha_s(K)<E_{\min} (K)$ holds.
\end{proof}

{\bf Proof of Theorem \ref{baho}}. The operator $V_\alpha$ defined by
\eqref{potential} has form $V_\alpha=\mu_{\alpha}V$, where
$$(Vf)(p)= (2\pi)^{-3}\int_{\T^3} f(s,p)ds .$$
One can check that
$$\sup_{||f||=1}(Vf,f)=1$$
and hence
$$(H(K)f,f)=(H_0(K)f,f)-\mu^0_\alpha(Vf,f)-\mu^0_\beta(Vf,f)-\mu_\gamma(Vf,f)$$$$
=(H_\alpha(K)f,f)-\mu^0_\beta(Vf,f)-\mu_\gamma(Vf,f)\geq
(H_\alpha(K)f,f)-(\mu^0_\beta+\mu_\gamma)\sup_{||f||=1}(Vf,f).$$

Thus
$$\inf_{||f||=1}
(H(K)f,f)\geq \inf_{||f||=1} (H_\alpha(K)f,f)
-\mu^0_\beta-\mu_\gamma.$$

The definition (6.3) of $\tau^\alpha_s(K)$ imply that
$$\tau^\alpha_s(K)-\mu^0_\beta-\mu_\gamma\leq
\inf_{||f||=1}(H(K)f,f)=\tau_s(K)<\tau_{ess}(K).$$ $\Box$

Let $W_\alpha(K,z),\alpha=1,2,3$ be the operators on
$L_2(({\T}^3)^2)$ defined as
$$
W_\alpha(K,z)=I+V_\alpha^{\frac{1}{2}}
R_\alpha(K,z)V_\alpha^{\frac{1}{2}},
 $$
where  $R_\alpha(K,z),\,\alpha=1,2,3$ are the resolvents of
$H_\alpha(K),\alpha=1,2,3.$ One can check that
$$
W_\alpha(K,z)=(I-V^{\frac{1}{2}}_\alpha
R_0(K,z)V^{\frac{1}{2}}_\alpha)^{-1},$$ where $R_0(K,z)$ the
resolvent of the operator $H_0(K).$

For $z < \tau_{ess}(K),\,\,\tau_{ess}(K)=\inf \sigma_{ess}(H(K))$
the operators $W_\alpha(K,z),\alpha=1,2,3$ are positive.

Denote by ${\bf \cL}= L^{(3)}_2(({\T}^3)^2)$ the space of vector
functions $w$ with components $w_{\alpha}\in L_2(({\T}^3)^2),
{\alpha}=1,2,3.$

Let
$$
 {\bf T}(K,z),\,  z \leq \tau_{ess}(K)
$$
be the operator on ${\cL} $ with the entries
\begin{align*}
&{\bf T}_{ \alpha\alpha } ( K, z) = 0, \\
&{\bf T}_{ \alpha\beta } ( K, z) =
W^{\frac{1}{2}}_{\alpha}(K,z)V_\alpha^{\frac{1}{2}}
R_0(K,z)V_\beta^{\frac{1}{2}}W^{\frac{1}{2}}_{\beta}(K,z).
\end{align*}
For any bounded self-adjoint operator $A$ acting in the Hilbert
space ${\cH}$ not having any essential spectrum on the right of
the point $z$ we denote by	${\cH}_A(z)$ the subspace such that
$(Af,f) > z(f,f)$ for any $f \in {\cH}_A(z)$ and set
$n(z,A)=\sup_{\cH_A(z)}\dim{\cH}_A(z)$. By the definition of
$N(K,z)$ we have
$$
N(K,z)=n(-z,-H(K)),\,-z > -\tau_{ess}(K).
$$
The following lemma is a realization of well known
Birman-Schwinger principle for the three-particle Schr\"{o}dinger
operators on lattice (see [23,25]).
\begin{lemma}\label{b-s}
 For $z<\tau_{ess}(K)$ the operator ${\bf T}(K,z)$ is
compact and continuous in $z$ and
$$
N(K,z)=n(1,{\bf T}(K,z)).
$$
\end{lemma}
\begin{proof} First verify the equality
\begin{equation}\label{tenglik}
N(K,z)=n(1,R^{\frac{1}{2}}_0(K,z)V_TR^{\frac{1}{2}}_0(K,z)),\quad
V_T=V_1+V_2+V_3.
\end{equation}
 Assume that $u \in
{\cH}_{-H(K)}(-z)$, that is, $((H_0(K)-z)u,u) < (V_Tu,u).$ Then $$
(y,y) < (R^{\frac{1}{2}}_0(K,z)V_TR^{\frac{1}{2}}_0(K,z)y,y),\quad
y=(H(K)-z)^{\frac{1}{2}}u.
$$
Thus $N(K,z) \leq
n(1,R^{\frac{1}{2}}_0(K,z)V_TR^{\frac{1}{2}}_0(K,z))$. Reversing
the argument we get the opposite inequality, which proves
\eqref{tenglik}.

Now we use the	following well known fact.
\begin{proposition}\label{propos}
Let $T_1,T_2$ be bounded operators. If $z\neq0$ is an eigenvalue
of $T_1T_2$ then $z$ is an eigenvalue for $T_2T_1$ as well of the
same algebraic and geometric multiplicities.
\end{proposition}
Using Proposition \ref{propos} we get
$$
n(1,R^{\frac{1}{2}}_0(K,z)V_TR^{\frac{1}{2}}_0(K,z))=n(1,{\bf
M}(K,z)),
$$
where ${\bf M}(K,z)$ the operator on ${\cL}$ with the entries

$$
M_{{\alpha \beta}}=V^{\frac{1}{2}}_\alpha
R_0(K,z)V^{\frac{1}{2}}_\beta,\quad \alpha,\beta =1,2,3.
$$

Let us check that
$$
n(1,{\bf M}(K,z))=n(1,{\bf T}(K,z)).
$$
We shall show that for any $u \in {\cH}_{{\bf M}(K,z)}(1)$ there
exists $y \in{\cH}_{{\bf T}(K,z)}(1)$ such that $(y,y)<({\bf
T}(K,z)y,y).$ Let $u \in {\cH}_{{\bf M}(K,z)}(1),$ 
that is,
$$
\sum_{\alpha=1}^3(u_\alpha,u_\alpha)<
\sum_{\alpha,\beta=1}^3(V^{\frac{1} {2}}_\alpha
R_0(K,z)V^{\frac{1}{2}}_\beta u_\beta,u_\alpha)
$$
and hence
$$
\sum_{\alpha=1}^3((I-V^{\frac{1}{2}}_\alpha R_0(K,z)V^{\frac{1}{2}
}_\alpha)u_\alpha,u_\alpha)< \sum_{\beta \neq \alpha
=1}^3(V^{\frac{1}{2} }_\alpha R_0(K,z)V^{\frac{1}{2}}_\beta
u_\beta,u_\alpha).
$$
Denoting by $y_\alpha=(I- V_{\alpha}^{\frac{1}{2}}R_0(K,z)V_\alpha
^{\frac{1}{2}})^{\frac{1}{2}} u_\alpha
$
we have
$$
\sum_{\alpha=1}^3(y_\alpha,y_\alpha)< \sum_{\beta \neq \alpha
=1}^3 (W^{\frac{1}{2}}_\alpha(K,z)V^{\frac{1}{2}}_\alpha R_0(K,z)
V^{\frac{1}{2}}_\beta W^{\frac{1}{2}}_\beta
(K,z)y_\beta,y_\alpha),
$$
that is,
$
(y,y)\leq ({\bf T}(K,z)y,y).
$
Thus
$
n(1,{\bf M}(K,z)) \leq n(1,{\bf T}(K,z)).
$

By the same way one can check
$
n(1,{\bf T}(K,z)) \leq n(1,{\bf M}(K,z)) .
$
\end{proof}

 Set
$$
\Sigma(K):= \cup_{\alpha=1}^{3} \sigma(H_\alpha(K)).
$$
 Denote	 by $ L^{ (3) } _ 2({\T}^3)$ the space of vector-functions $w=(w_1,w_2,w_3)$,
 $w_{\alpha}\in L_2({\T}^3), {\alpha}=1,2,3$
and define	compact operator $T( K, z),	 z \in {\bf C} \setminus
\Sigma(K)$ on $L^{(3)}_2({\T}^3) $ with the entries
\begin{align*}
&T_{ \alpha\alpha } ( K, z) = 0,\\
&( T_{\alpha\beta}(K,z)w_\beta)(p_\alpha)=
\sqrt{\mu_\alpha\mu_\beta}(2\pi)^{-3}
 \int\limits_{{\T}^3} \frac{\Delta^{-\frac{1}{2}}_\alpha(K, p_\alpha, z)
\Delta^{-\frac{1}{2}}_\beta (K, p_{\beta}, z)
}{E_{\alpha\beta}(K;p_\alpha,p_{\beta})-z}w_\beta(p_{\beta})
d p_{\beta},\\
&w \in L^{(3)}_2({\T}^3),
\end{align*}  where
\begin{equation} \Delta_\alpha (K,
p, z ):=\Delta_\alpha ((m_\beta+m_\gamma) K-p,
z-\varepsilon_\alpha (m_{\alpha}K-p)).
\end{equation}
Now we show that the numbers of eigenvalues greater than $1$ of
the operators ${\bf T}(K,z)$ and $T(K,z)$ are coincide.

Let $$\Psi=\text{diag}\{\Psi_1,\Psi_2,\Psi_3\}:L^{(3)}_
2(({\T}^3)^2)\to L^{(3)}_ 2({\T}^3)$$ be the operator with the
entries $$(\Psi_\alpha f)(p_\alpha)={(2\pi)^{-\frac{3}{2}}}
\int_{\T^3}f(q,p_\alpha)dq,\quad \alpha=1,2,3
$$ and
$\Psi^*=\text{diag}\{\Psi^*_1,\Psi^*_2,\Psi^*_3\}$ its adjoint.

\begin{lemma}\label{fadd}
The following equalities $${\bf T}(K,z)=\Psi^*T(K,z)\Psi
\quad\text{and}\quad n(1,{\bf T}(K,z))=n(1,{T}(K,z))$$ hold.
\end{lemma}
\begin{proof}
One can easily check that the equalities
\begin{equation}\label{faddeev}\Psi_\alpha f={(2\pi)^{\frac{3}{2}}}\mu^{-\frac{1}{2}}_\alpha
V^{1/2}_\alpha f \quad\text{and}\quad W^{1/2}_\alpha
V^{1/2}_\alpha f= \Delta_\alpha^{-\frac{1}{2}}(K,p_\alpha,z)
V^{1/2}_\alpha f
\end{equation} hold. The equalities \eqref{faddeev} implies the
first equality of Lemma \ref{fadd}. By Proposition \ref{propos} we
have $$n(1,{\bf
T}(K,z))=n(1,\Psi^*T(K,z)\Psi)=n(1,{T}(K,z)\Psi\Psi^*)=n(1,{T}(K,z).$$
\end{proof}

  Now we establish a location of the essential
spectrum of $H(K).$ For any $K\in {\T}^3$ and $z \in {\bf C}
\setminus \Sigma(K) $ the kernels of the operators
$T_{\alpha\beta}(K,z),\alpha,\beta=1,2,3$ are continuous functions
on $({\T}^3)^2$. Therefore the Fredholm determinant $D_{K}(z)$ of
the operator $I- T( K,z)$, where $I$ is the identity operator in
$L^{(3)}_2({\T}^3),$ exists and is a real-analytic function on ${\bf C}
\setminus {\Sigma(K)}.$ The
 following theorem is a lattice analog of the well known
Faddeev's result for the three-particle Schr\"{o}dinger operators
with the zero-range interactions and can be prove similarly	 to
 that of the identical particle case (see [12]).
\begin{theorem}
For any $K \in {\T}^3$ the number $z \in {\bf C} \setminus
\Sigma(K)$ is an eigenvalue of the operator $H(K)$ if and only if
the number 1 is eigenvalue of ${T}(K,z).$
\end{theorem}$\Box$

 According to  Fredholm's theorem the following lemma
holds.
\begin{lemma}\label{nol}The number $z \in {\bf C} \setminus {\Sigma(K)}$
is an eigenvalue of the operator $H(K)$
 if and only if
$$
D_{K}(z)=0.
$$
\end{lemma} $\Box$

{\bf Proof of Theorem} \ref{esss}. By the definition of the
essential spectrum, it is easy to show that $ \Sigma(K) \subset
{\sigma}_{ess}(H(K)).$ Since the function $D_K(z)$ is analytic in
${\bf C}\setminus \Sigma(K)$ by Lemma \ref{nol} we conclude that
the set
$$
\sigma(H(K))\setminus \Sigma(K)=\{z:\, D_K(z)=0\}
$$
is discrete. Thus
  $$
  \sigma(H(K))\setminus \Sigma(K)
  \subset  \sigma(H(K))\setminus{\sigma}_{ess}(H(K)).
  $$
Therefore the inclusion ${\sigma}_{ess}(H(K))\subset \Sigma(K)$
holds. $\Box$

Now we are going to proof the finiteness of $N(K,\tau_{ess}(K))$
for $K\in U_\delta^0(0),\delta>0$ sufficiently small. First we
shall prove that the operator $T(K,\tau_{ess}(K))$ belongs to the
Hilbert-Schmidt class.

The point $p=0$ is the non degenerate minimum of the functions
$\varepsilon_\alpha(p)$ and $z_\alpha(p)$ (see \eqref{corol}) and
hence $p=0$ is the non degenerate  minimum of $Z_\alpha(0,p)$
defined by $$ Z_{\min
}^{(\alpha)}(K,p):=\varepsilon_{\alpha}(m_{\alpha} K-p)+ z_\alpha
((m_\beta+m_\gamma)K+p).$$
 Simple computations gives
$$
 \big ( \frac{\partial^2 Z_\alpha(0,0)}{\partial p^{(i)} \partial p^{(j)}}
 \big )_{i,j=1}^3=
\frac{l_1l_2+l_2l_3+l_1l_3} {2(l_\beta + l_\gamma)} \left (
\begin{array}{lll}
1\,\,0\,\,0\\
0\,\,1\,\,0\\
0\,\,0\,\,1
\end{array}
\right ).
$$
Therefore for all $K \in U_\delta^0(0)$ at the non degenerate
minimum	 point $p^Z_{\alpha}(K)\in U_\delta^0(0)$ of the function
$Z_\alpha(K,p),\,K \in U_\delta^0(0)$ the inequality
$$
B(K)=\big (\frac{dZ^2_\alpha}{dp^{(i)} dp^{(j)}} (K,p^Z_\alpha(K))
 \big )_{i,j=1}^3 >0
$$
holds. Hence   the asymptotics \be\label{Z}
Z_\alpha(K,p)=\tau^\alpha_s(K)+
(B(K)(p-p^Z_\alpha(K)),p-p^Z_\alpha(K))
+o(|p-p^Z_\alpha(K)|^2)\,\,\mbox{as}\,\,|p-p^Z_\alpha(K)| \to 0
\ee is valid, where
$\tau^\alpha_s(K)=Z_\alpha(K,p^Z_{\alpha}(K)).$ From Lemma
\ref{tasvir} we conclude that for all $K\in U_\delta^0(0),p\in
U_{\delta(K)}(p^Z_\alpha(K))$ the equality
\begin{equation}\label{nondeger}
\Delta_\alpha(K,p,\tau^\alpha_{s}(K))
=(Z_\alpha(K,p)-\tau^\alpha_{s}(K)) \hat
\Delta_\alpha(K,p,\tau^\alpha_{s}(K))
 \end{equation}
holds, where
$\hat
\Delta_\alpha(K,p^Z_\alpha(K),\tau^\alpha_{s}(K))\neq 0.$
Putting \eqref{Z} into \eqref{nondeger} we get	the following
\begin{lemma} Let$\mu_\alpha=\mu^0_\alpha,\alpha=1,2,3.$ Then
  for any $K \in U^0_\delta(0),\delta=\delta(K)$ sufficiently small, there are
  positive nonzero
constants $c$ and $C$ depending on $K$ and
$U_{\delta(K)}(p^Z_\alpha(K))$ such that  for all $p\in
U_{\delta(K)}(p^Z_\alpha(K))$ the following inequalities
 \begin{equation}\label{otsenka2}
 c|p -p^Z_\alpha(K)|^2 \leq
\Delta_\alpha(K,p,\tau^{\alpha}_{s}(K))\leq C|p -p^Z_\alpha(K)|^2
\end{equation}
hold.
\end{lemma}
$\Box$
\begin {lemma}\label{G-S}
Let $\mu_\alpha \leq \mu_\alpha^0$ for all $\alpha=1,2,3$.
 Then for any $K\in
U_\delta^0(0),\delta>0$ sufficiently small, the operator $T(K,
\tau_{ess}(K))$ belongs to the Hilbert-Schmidt class.
\end{lemma}
\begin{proof} As we shall see that it is sufficient to prove Lemma \ref{G-S}
 in the case $\mu_\alpha =\mu_\alpha^0$ for all $\alpha=1,2,3$.
 By Lemma \ref{inequality} we have
\begin{equation}\label{inequal 2} \tau_{ess}(K)=\min_{\alpha}\tau^\alpha_s(K) <
E_{\min}(K),\,K \in U_\delta^0(0).
\end{equation}
The operator $h_{\alpha}(0)$ has a zero energy resonance. By
Theorem \ref{mavjud} the operator $h_{\alpha}(k),\,k\in \T^3,\, k\neq
0$ has a unique eigenvalue $z_\alpha(k),$
$z_\alpha(k)<E_{\min}^{(\alpha)}(k).$

Since $ \tau^\alpha_s(K)= \min_{p \in {\T}^3}Z_\alpha(K,p)$ the
function $Z_\alpha(K,p)$ has a unique minimum and hence for all $
p\in \T^3\setminus U_\delta(p^Z_\alpha(K))$ we obtain
\begin{equation}\label{otsenka}
\Delta_\alpha(K,p,\tau^\alpha_{s}(K))\geq C>0.
\end{equation}
According to \eqref{inequal 2} for all $p_\alpha,p_\beta\in {\T}^3$
and $K\in U_\delta^0(0)$ the inequality
\begin{equation}\label{inequal3}
E_{\alpha\beta}(K;p_\alpha,p_\beta) -\tau^\alpha_{s}(K)\geq
E_{\min}(K)-\tau^\alpha_{s}(K)>0
\end{equation} holds.
Using \eqref{otsenka2}, \eqref{otsenka} and taking into account
\eqref{inequal3} we can make certain that for all $K\in
U_\delta^0(0)$ and $p_\alpha \in U_\delta(p^Z_\alpha(K)),\,
p_\beta \in U_\delta(p^Z_\beta(K))$
the modules of the kernels $T_{\alpha\beta}
(K,\tau_{ess}(K);p_\alpha,p_\beta)$ of the integral operators
$T_{\alpha\beta} (K,\tau_{ess} (K))$ can be estimated by
$$
\frac{C_0(K)}{|p_\alpha-p^Z_\alpha(K)||p_\beta-p^Z_\beta(K)|}+C_1,
$$ where $C_0(K)$ and $C_1$ some constants.
Taking into account \eqref{inequal 2} we conclude that
$$T_{\alpha\beta} (K,\tau_{ess}(K)),\,\alpha,\beta=1,2,3$$ are
Hilbert-Schmidt operators. Thus, $T(K,\tau_{ess}(K))$ belongs to
the Hilbert-Schmidt class.
\end{proof}
Now we shall prove the finiteness of $N(K, \tau_{ess}(K))$
\eqref{finite} and a generalization of the Birman-Schwinger
principle for  three-particle discrete Schr\"{o}dinger
operators on lattices.

\begin{theorem}\label{g.b-s} Assume Hypothesis \ref{hypoth}.Then for the number $N(K,
\tau_{ess}(K))$ the relations
$$
 N(K, \tau_{ess}(K))=n(1,T(K, \tau_{ess}(K)))\leq
 \lim_{\gamma\to 0}n(1-\gamma,T(K,\tau_{ess} (K)))
$$
hold.\end{theorem}

 \begin{proof}	By Lemmas \ref{b-s} and \ref{fadd} we have
$$
N(K,z)=n(1,T(K,z))\,\,\mbox{as}\,\,z<\tau_{ess}(K)
$$
and by Lemma \ref{G-S} for any $\gamma\in [0,1)$ the number
$n(1-\gamma,T(K,\tau_{ess}(K))),\,K \in U_\delta^{o}(0) $ is
finite. Then according to the Weyl inequality
$$n(\lambda_1+\lambda_2,A_1+A_2)\leq
n(\lambda_1,A_1)+n(\lambda_2,A_2)$$ for all $z<\tau_{ess}(K)$ and
$\gamma \in (0,1)$ we have
$$
N(K,z)=n(1,T(K,z))\leq
n(1-\gamma,T(K,\tau_{ess}(K)))+n(\gamma,T(K,z)-T(K,\tau_{ess}(K))).
$$
Since $T(K,z)$ is continuous from the left up to
$z=\tau_{ess}(K),\,K\in U_\delta^0(0)$, we obtain
$$
\lim_{z\to \tau_{ess}(K)} N(K,z)= N(K,\tau_{ess}(K))\leq
n(1-\gamma,T(K,\tau_{ess}(K)))\,\, \mbox{for all}\,\, \gamma \in
(0,1)
$$
and so $$N(K,\tau_{ess}(K))\leq \lim _{\gamma\to
0}n(1-\gamma,T(K,\tau_{ess}(K))).$$ By definition of $n(1,T(K,z))$
for any $\gamma \in (0,1)$ the equality
$$n(1-\gamma,T(K,\tau_{ess}(K)))\geq n(1,T(K,\tau_{ess}(K)))$$ holds.
Since $N(K,\tau_{ess}(K))$ is finite we have
$N(K,\tau_{ess}(K)-\gamma)=N(K,\tau_{ess}(K))$ for all small
enough $\gamma \in (0,1)$. Therefore using Lemma \ref{b-s} and
continuity of $N(K,z)$ from the left we derive the equality
$$n(1,T(K,\tau_{ess}(K)))=\lim _{\gamma\to 0}n(1,T(K,\tau_{ess}(K)-\gamma))$$$$=
\lim _{\gamma\to 0}N(K,\tau_{ess}(K)-\gamma)=N(K,\tau_{ess}(K)).$$
\end{proof}

\section{Asymptotics for the number of
eigenvalues of the	operator $H(K)$}

We recall that in this section we closely follow  A.Sobolev's method
to derive the asymptotics for the number of eigenvalues of
$H(K)$ (Theorem \ref{asym.zK}).
 As we shall see, the discrete
spectrum asymptotics of the operator $T(K,z)$ as $|K|\to 0$ and
$z\to -0$ is determined	 by the integral operator
$${\bf S}_{{\bf
r}},\,{\bf r}=1/2 | \log (\frac{|K|^2}{2M}+|z|)|$$ in
$$L_2((0,{\bf r})\times {\bf \sigma}^{(3)}),\, {\bf
\sigma}=L_2({\bf S}^2), $$ with the kernel
$S_{\alpha\beta}(x-x';<\xi, \eta>),\,\xi, \eta \in {\bf S}^2,{\bf
S}^2$ is unit sphere in $\R^3,$ where

\begin{equation}\label{sobolev}
 S_{\alpha\alpha}(x;t)=0,\quad
 S_{\alpha\beta}(x;t)=(2\pi)^{-2}\frac{u_{\alpha\beta}}
{\cosh(x+r_{\alpha \beta})+s_{\alpha\beta}t}
\end{equation}
and
$$
u_{\alpha\beta}=k_{\alpha\beta} \big(  \frac{l_{\beta\gamma} l_{\alpha\gamma}%
} {n_\alpha n_\beta}\big)^{\frac{1}{4}},\,
r_{\alpha\beta}=\frac{1}{2} \log
\frac{l_{\alpha\gamma}}{l_{\beta\gamma}} ,\, s_{\alpha\beta}=\frac{l_\gamma%
} {(l_{\alpha\gamma} l_{\beta\gamma})^{\frac{1}{2}}},
$$
$k_{\alpha\beta}$ being such that $k_{\alpha\beta}=1$ if both
subsystems $\alpha$ and $\beta$ have zero resonances, otherwise $
k_{\alpha\beta}=0$. The eigenvalues asymptotics for the operator
${\bf S}_{\bf r}$ has been
 studied in detail by Sobolev [23], by employing an argument used
 in the
calculation of the canonical distribution of Toeplitz operators.
We here summarize some results obtained in[23].
\begin{lemma} The following equality
$$
\lim\limits_{{\bf r}\to \infty} \frac{1}{2}{\bf r}^{-1}n(\lambda,{\bf
S}_{\bf r})= {\cU(\lambda)}
$$
holds, where the function ${\cU(\lambda)}$ is continuous in
 $\lambda>0$ and ${\cU}_0$ in \eqref{asym.K} defined as ${\cU}_0=\cU(1)$.
\end{lemma}

\begin{lemma}\label{comp.pert} Let $A (z)=A_0 (z)+A_1 (z),$ where $A_0$ $(A_1)$ is
compact and continuous in $z<0$ $(z\leq 0).$  Assume that for some
function $f(\cdot),\,\, f(z)\to 0,\,\, z\to -0$ the limit
$$
\lim_{z\rightarrow -0}f(z)n(\lambda,A_0 (z))=l(\lambda),
$$
exists and is continuous in $\lambda>0.$ Then the same limit
exists for $A(z)$ and
$$
\lim_{z\rightarrow -0}f(z)n(\lambda,A (z))=l(\lambda).
$$
\end{lemma}$\Box$

Now we are going to reduce the study of the asymptotics for the operator
$T(K,z)$ to that of the asymptotics of ${\bf S}_{\bf r}.$

From definition of the functions $\varepsilon_\alpha$  and
$E_{\min}^{(\alpha)}$  (see \eqref{min}) we obtain that
$$\varepsilon_\alpha(p)=\frac{l_\alpha}{2}p^2+O(|p|^4)\,\,\mbox{as}\,\,p\to
0$$ and $$E_{\min}^{(\alpha)}(k)= \frac{1}{2}\frac{l_\beta
l_\gamma} {l_\beta +l_\gamma}|k|^2+O(|k|^4)\,\,\mbox{as}\,\,k\to
0,$$
\begin{equation}\label{asymp1} E_{\alpha \beta}(K;
p,q)= \frac{(l_\alpha + l_\gamma)}{2} p^2+	l_\gamma (p,q)+
\frac{(l_\beta +l_\gamma)}{2}q^2 + \frac{K^2}{2M}+ O(|K|^4
+|p|^4+|q|^4)\,\, \text{as}\,\, K,p,q\rightarrow 0.
\end{equation}
From Lemma \ref{razlojeniya} we easily receive	the following
\begin{lemma} For any $K
\in U_{\delta}(0)$	and $z \in [-\delta,0]$ we have
\begin{equation}\label{asymp2} \Delta_\alpha(K,p, z) =
\frac{\mu^0_{\alpha}} {2 \pi(l_\beta +l_\gamma)^{\frac{3}{2}}}
\left [ n_\alpha p^2 +\frac{K^2}{M}-2 z
 \right ]^{\frac{1}{2}}+
 O(|K|^2+|p|^2+|z|)\,\, \text{as}\,\, K,p,z\rightarrow 0,
 \end{equation}
 where $$ n_\alpha\equiv\frac{l_1l_2+l_1l_3+l_2l_3}
 {l_\beta
+l_\gamma}.$$\end{lemma}

The following theorem is basic for the proof of the asymptotics
\eqref{asym.K}.
\begin{theorem}\label{main} The equality
$$
\lim\limits_{\frac{|K|^2}{M}+|z|\to 0} \frac{n(1,T(K,z))}
{|log(\frac{|K|^2}{M}+|z|)|} =\lim\limits_{{\bf r}\to \infty}
\frac{1}{2}{\bf r}^{-1}n(1,{\bf S}_{\bf r})
$$
holds.
\end{theorem}

\begin{remark} Since $\cU(.)$ is continuous in $\lambda,$ according
to Lemma \ref{comp.pert} a
compact and continuous up to $z=0$ perturbations of the operator
$A_0(z),$ do not contribute to the asymptotics \eqref{asym.K}.
During the proof of Theorem \ref{main} we use this fact without
further comments. First we prove  Theorem \ref{main} under the
condition that all two-particle operators have zero energy
resonances, that is, in the case where
$\mu_1=\mu_1^{0},\mu_2=\mu_2^{0}$ and $\mu_3=\mu_3^{0}.$ The case
where only two operators $h_\alpha(0)$ and $h_\beta(0)$ have zero
energy resonance can be proven similarly.
\end{remark}
{\bf Proof of Theorem.}\ref{main} Let
$T(\delta,\frac{K^2}{2M}+|z|)$ be an operator on
$L_2^{(3)}({\T}^3)$ with the entries
\begin{align*}
&T_{\alpha \alpha}(\delta;\frac{K^2}{2M}+|z|)=0,\\
&(T _{\alpha \beta}(\delta;\frac{K^2}{2M}+|z|)w)(p)\\ &=
D_{\alpha\beta} \int\limits_{\T^3} \frac{ \chi_\delta (p)
\chi_\delta (q) (n_\alpha p^2+ 2(\frac{K^2}{2M}+|z|))^{-1/4}
(n_\beta q^2+ 2(\frac{K^2}{2M}+|z|))^ {-1/4} } {l_{\beta
\gamma}q^2+2l_{\gamma}(p,q)+l_{\alpha\gamma}p^2
 +2(\frac{K^2}{2M}+|z|)} w(q)d q,
\end{align*}
where $$ D_{\alpha\beta}=\frac{l_{\alpha \gamma}^{\frac{3}{4}}
l_{\beta\gamma}^{\frac{3}{4}}} {2 \pi^2},\,\,\alpha ,\beta,\gamma
=1,2,3
 $$
and $\chi_\delta(\cdot)$ is the	 characteristic function of
$U_\delta(0)=\{ p:\,\, |p|<\delta \}.$
\begin{lemma}  The operator $T(K,z)-T(\delta;
\frac{K^2}{2M}+|z|)$ belongs to the Hilbert-Schmidt class and is
continuous in $K\in U_\delta (0)$ and $z\leq 0.$
\end{lemma}
\begin{proof}

Applying  asymptotics \eqref{asymp1} and \eqref{asymp2} one can
estimate the kernel of the operator $T_{\alpha\beta}
(K,z)-T_{\alpha \beta}(\delta; \frac{K^2}{2M}+|z|)$ by
\be
 C [ (p^2+q^2)^{-1} +
|p|^{-\frac{1}{2}}(p^2+q^2)^{-1} +
(|q|^{-\frac{1}{2}}(p^2+q^2)^{-1}+1 ] \ee and hence the operator
$T_{\alpha\beta} (K,z)-T_{\alpha \beta}(\delta;
\frac{K^2}{2M}+|z|)$
 belongs to the Hilbert-Schmidt class for all $K\in U_{\delta}(0)$ and
$z \leq 0.$ In combination with the continuity of the kernel of
the operator in $K\in U_\delta (0)$ and $z<0$ this	gives	the
continuity of $T(K,z)-T(\delta;\frac{K^2}{2M}+|z|)$ in $K\in
U_\delta (0)$ and $z\leq 0.$
\end{proof}
 The space of vector-functions
$w=(w_1,w_2,w_3)$ with coordinates having support in $U_\delta(0)$
 is an invariant subspace
for the operator $T(\delta,\frac{K^2}{2M}+|z|).$

Denote by ${\cL}_\delta$ the space of vector-functions $w =
(w_1,w_2,w_3),\,\, w_{\alpha} {\in} L_2(U_{\delta}(0)), $ that is,
$${\cL}_\delta ={\oplus}_{\alpha=1}^3 L_2(U_{\delta}(0)).$$
Let	 $T_0(\delta,\frac{K^2}{2M}+|z|)$ be the restriction of the
operator $T (\delta,\frac{K^2}{2M}+|z|)$ to the invariant subspace
${\cL}_\delta.$ One	 verifies  that the operator
$T_0(\delta,\frac{K^2}{2M}+|z|)$ is unitarily equivalent to the
operator $T_1(\delta,\frac{K^2}{2M}+|z|)$ with entries
\begin{align*}
&T^{(1)}_{\alpha\alpha}({\delta};\frac{K^2}{2M}+|z|)=0,\\
&(T^{(1)}_{\alpha\beta}({\delta};\frac{K^2}{2M}+|z|)w)(p)=
D_{\alpha\beta}\int\limits_{U_{r}(0)} \frac{(n_\alpha p^2+2 )^{-1/4}
(n_\beta q^2+2)^{-1/4}}{ l_{\beta
\gamma}q^2+2l_{\gamma}(p,q)+l_{\alpha\gamma}p^2+2} w(q)dq &
\end{align*}
acting in $L_2^{(3)}(U_r(0)),\,\,
r=(\frac{|K|^2}{2M}+|z|)^{-\frac{1}{2}}.$ The equivalence is
performed by the unitary dilation
 $${\bf B}_r=diag\{B_r,B_r,B_r\}:L_2^{(3)}
(U_\delta(0)) \to L_2^{(3)}(U_r(0)),\quad
 (B_r f)(p)=(\frac{r}{\delta})^{-3/2}f(\frac{\delta}
{r}p).$$ Further, we may replace $$(n_\alpha p^2+2)^{-1/4},\,
(n_\beta q^2+2)^{-1/4} \quad \mbox{ and}\quad
l_{\beta\gamma}q^2+2l_{\gamma}(p,q)+l_{\alpha\gamma}p^2+2$$
 by
$$(n_\alpha p^2)^{-1/4}(1-\chi_1(p)),\,\, (n_\beta
q^2)^{-1/4}(1-\chi_1(q))
 \quad \mbox{ and}\quad
l_{\beta\gamma}q^2+2l_{\gamma}(p,q)+l_{\alpha\gamma}p^2,$$
  respectively, since the error will be
a Hilbert-Schmidt operator continuous up to $K=0$ and $z=0$. Then
we get the operator $T^{(2)}(r)$ in $L_2^{(3)}(U_r(0) \setminus
U_1(0))$ with entries
\begin{align*}
&T^{(2)}_{\alpha\alpha}(r)=0,\\
&(T^{(2)}_{\alpha\beta}(r)w)(p)= (n_\alpha n_\beta)^{-\frac{1}{4}}
D_{\alpha\beta}\int\limits_{U_r(0)\setminus U_1(0)}
\frac{|p|^{-1/2}| q|^{-1/2}} {l_{\beta
\gamma}q^2+2l_{\gamma}(p,q)+l_{\alpha\gamma}p^2} w(q)dq.
\end{align*}

This operator $T^{(2)}(r)$ is unitarily equivalent to the integral
operator ${\bf S}_{{\bf r}}$
with entries \eqref{sobolev}. The equivalence is performed by the
unitary operator ${\bf M}=diag\{M,M,M\}:L_2^{(3)}(U_r(0) \setminus
U_1(0)) \longrightarrow L_2((0,{\bf r})\times {\bf
\sigma}^{(3)}),$ where $(M\,f)(x,w)=e^{3x/2}f(e^{ x}w), x\in
(0,{\bf r}),\, w \in {\bf S}^2.$

{\bf Acknowledgement} The authors grateful to Prof. R.A.Minlos and
Dr. J.I.Abdullaev for useful discussions, Prof. G.M.Graf
 and the referee for
useful critical remarks.

This work was supported by the DFG 436 USB 113/3 and DFG 436 USB
113/4 projects and the Fundamental Science Foundation of
Uzbekistan. The last two named authors gratefully acknowledge the
hospitality of the Institute of Applied Mathematics and of the
IZKS of the University Bonn. \end{document}